# Pixelation with concentration-encoded effective photons for molecular optical sectioning microscopy

Geng Wang[1,2]†, Rishyashring R. Iyer[1,2]†, Janet E. Sorrells[1,3]†, Edita Aksamitiene[1], Eric J. Chaney[1], Carlos A. Renteria[1,3], Jaena Park[1,3], Jindou Shi[1,2], Yi Sun[1,2], Stephen A. Boppart[1,2,3], Haohua Tu[1,2]*

[1]Beckman Institute for Advanced Science and Technology, University of Illinois at Urbana-Champaign, Urbana, IL 61801, USA

[2]Department of Electrical and Computer Engineering, University of Illinois at Urbana-Champaign, Urbana, IL 61801, USA

[3]Department of Bioengineering, University of Illinois at Urbana-Champaign, Urbana, IL 61801, USA

†These authors contributed equally to this work.

*Corresponding author. Emails: htu@illinois.edu

**Abstract**

Quality control in molecular optical sectioning microscopy is indispensable for transforming acquired digital images from qualitative descriptions to quantitative data. Although numerous tools, metrics, and phantoms have been developed, accurate quantitative comparisons of data from different microscopy systems with diverse acquisition conditions remains a challenge. Here, we develop a simple tool based on an absolute measurement of bulk fluorophore solutions with related Poisson photon statistics, to overcome this obstacle. Demonstrated in a prototypical multiphoton microscope, our tool unifies the unit of pixelated measurement to enable objective comparison of imaging performance across different modalities, microscopes, components/settings, and molecular targets. The application of this tool in live specimens identifies an attractive methodology for quantitative imaging, which rapidly acquires low signal-to-noise frames with either gentle illumination or low-concentration fluorescence labeling.

**Introduction**

With the call for public biological image archives[1] such as Image Data Resource[2], increasing awareness has been raised to improve the quantification[3] and reproducibility[4] of molecular (e.g., fluorescence) optical sectioning microscopy, including laser-scanning confocal/multiphoton microscopy and light-sheet microscopy. These aspects require rigorous quality control that can be loosely divided into 15 relatively independent subtasks/goals (Supplementary Table 1)[3-5]. One systematic effort has performed 6 of these subtasks (Supplementary Fig. 1) for wide-field and confocal laser-scanning microscopy platforms across imaging core facilities[6] but has not engaged the other 9 subtasks. More importantly, the corresponding tools, metrics, phantoms, and protocols would overburden a portable imaging facility, which requires one comprehensive calibration per travel to image intrinsically non-transportable living specimens[7]. The lack of a simple quality-control tool has hindered the widespread application of portable label-free laser-scanning multiphoton microscopy in dermatology[8] and neurosurgery[9] and could limit the ongoing dissemination of other modalities such as the traveling light-sheet wide-field fluorescence microscopy among biological laboratories (Flamingo project)[10].

Numerous designs of various live-cell imaging modalities[11] have challenged the development of a general quality-control tool. In this study, we propose a universal solution of quality control that engages all 15 subtasks with minimal calibration procedures, based on the simultaneous label-free auto-fluorescence multi-harmonic microscopy (SLAM) that integrates 4 modalities of two- and three-photon excited fluorescence and harmonics (2PF, SHG, 3PF, and THG)[12]. We upgrade this system to a portable system[13,14] (pSLAM) and an extended version (eSLAM) that incorporates a stabilized (>2000 hr) fiber supercontinuum source[15] (Supplementary Fig. 2). The corresponding 'label-free' aspect not only renders 2 sample-dependent subtasks irrelevant, but also mitigates plausible phototoxicity during time-lapse imaging[14] by inline monitoring the intrinsic phototoxicity indicator of elevated auto-fluorescence[16], which has been linked to impaired cell cloning[17]. Also, the use of multiphoton illumination ensures negligible out-of-focus background, while the resulting simultaneous multicolor detection at single-band excitation ensures aberration-free chromatic co-registration[12]. In this way, only 10 subtasks remain relevant (Supplementary Table 1).

With the deterministic (coherent) spectral broadening in a single-mode fiber[14,15,18] that guarantees stable illumination via laser-microscope alignment decoupling[19] (Fig. 1), we realize this elusive tool using diverse elements of photo-detection not directly related to imaging quality control[20-27] (Supplementary Table 2). Our tool, termed as Pixelating with Concentration-encoded Effective Photons (PCEP), integrates 3 independent subtasks to 'image' fluorophore solutions with known concentrations[28,29], while makes the other 7 subtasks dependent or feasible (Supplementary Table 1). This leads to a surprisingly simple procedure to monitor hardware failure or aging in SLAM-based imaging. Also, PCEP is validated by other forms of molecular optical sectioning microscopy with point-like (Supplementary Table 3) or camera-like photo-detection (see below), implying its broad applicability to alternative designs (Supplementary Fig. 3). This may unify often proprietary image pixel representations from different microscopy vendors with a unit of absolute measurement (effective photon) directly related to the local concentration of a (labeled) biomolecule of interest.

**Results**

*In situ absolute measurement using stable illumination*

Typically, the performance of a photodetector is either measured without placing it *in situ* (in a pre-aligned microscope with fixed optical components and alignments) or simply taken as original factory calibration without measuring over time, both of which are unsuitable for monitoring plausible failure or aging. It is thus an important progress to measure a photomultiplier tube (PMT) *in situ*, using homogeneous samples such as the solutions of reduced nicotinamide adenine dinucleotide (NADH)[20] and flavin adenine dinucleotide (FAD), i.e., well-known intrinsic fluorophores in cellular metabolism. A similar experiment has allowed absolute measurement of multiphoton excitation molecular cross-section[21], suggesting the feasibility to correlate (encode) the concentration of a fluorophore in solution with the number of detected photons. Motivated by these studies, we aimed to assess analog photo-detection performance by a simple *in situ* absolute measurement, using the illumination from a fiber supercontinuum laser with stable beam pointing, spatial mode propagation, and spectral power[15].

Analog photo-detection noise consists of Poisson noise that includes the shot noise and excess noise[22] (i.e., multiplicative noise), and non-Poisson noise that includes the additive noise[23]. Thus, the Poisson-noise-dominated dynamic range (PDR) of a point-like analog photodetector can be determined experimentally wherever the non-Poisson noise is negligible. For detected fluorescence photons from a fluorophore solution within the PDR, the *in situ* measured signal-to-noise-ratio (SNR) in a small/flat-field area (e.g., several square micrometers[20]), i.e., mean versus standard deviation (STD) of the pixelated arbitrary intensity value from PMT analog output, satisfies the Poisson statistics of

$$\text{SNR}|_{\text{theory}} = \frac{m\bar{N} \text{ (signal)}}{\sqrt{m\bar{N}} \text{ (noise)}} = \sqrt{m\bar{N}} \text{ (where } \bar{N} = \frac{N_D}{1+\varepsilon}\text{)} \qquad \text{Eq. 1}$$

$$\text{SNR}|_{\text{experiment}} = \frac{\text{Mean}}{\text{STD}}\bigg|_m = \sqrt{f_C C} \qquad \text{Eq. 2}$$

$$\text{SNR}|_{\text{experiment}} = \frac{\text{Mean}}{\text{STD}}\bigg|_m = \sqrt{f_P P^n} \qquad \text{Eq. 3}$$

where $\overline{N_D}$ is the average signal/fluorescence photons detected by the PMT per pulse (or per excitation cycle to incorporate linear optical microscopy), $\varepsilon$ is detector-dependent excess noise factor which is a constant for an analog PMT (20-70%)[22] or zero for a photon-counting PMT (similar classification is applicable to array- or camera-like detectors), $\bar{N}$ is the corresponding effectively detected photons with a unit of "effective photon" at equal mean and STD, $m$ is the number of pulses per pixel within one frame or over multiple frames[24] in one pixelated measurement, $C$ is the concentration of a fluorophore of interest with a fitting parameter $f_C$ to attain $m\bar{N}$ (Eq. 1 and Eq. 2), and $n$ is the order of optical nonlinear process at average power $P$ while $f_P$ is the corresponding fitting parameter (Eq. 1 and Eq. 3).

Experimentally, we focused the illumination at a shallow depth (15±5 μm) inside the solutions with a small field-of-view (FOV) of <10×10 μm², i.e., a high-zoom raster scanning through a microscope objective (Fig. 1). By varying $C$ of NADH solutions at a constant $P$ or by varying $P$ on a NADH/FAD solution with constant $C$, we tested these equations across three laser-scanning multiphoton microscopes operated at either one pulse per pixel[24] or hundreds of pulses per pixel (Supplementary Table 3), so that each experimental point of SNR involved ≥9000 pulses. The parameter $f_C$ (or $f_P$) can be obtained from the single-parameter linear fit between experimental (mean/STD)² and $C$ (or $P^n$) before signal saturation according to Eq. 2 (or Eq. 3), as shown in Supplementary Fig. 4 (or Supplementary Fig. 5). Indeed, the theoretical log-scale linear relation between SNR and $\bar{N}$ (Eq. 1) was validated by the corresponding experimental lines (Eq. 2 or Eq. 3) indicative of PDRs (Fig. 2). This was valid despite the differences in microscope, $n$ (2 or 3), $m$ (1-320), variable/parameter ($C/f_C$ or $P^n/f_P$), fluorophore (NADH or FAD), photodetector (regular or hybrid PMT), and the temporal window of signal collection optimized[14] for the largest PDR (Supplementary Fig. 6). Without detecting any effect from somewhat variable imaging depth (Fig. 1), we obtained the same effective photons from a given $C$ at two PMT gains that produced 10-fold different arbitrary intensity values (Fig. 2a, arrowheads). Thus, for fixed components and settings, the information of $C$ was encoded by the effective photons from an absolute measurement, not the gain-dependent arbitrary intensity values.

These results suggest that Poisson noise-limited performance may be ensured over time by obtaining the same upper limit of a PDR at a certain gain (Fig. 2a,c) and the same lower limit of the PDR at a high gain (Fig. 2b,d). The PDR with variable $P$ (Fig. 2c, left) is often understood through the shot-noise-limited photon transfer curve (PTC) (Fig. 2c, right), which has assessed PMTs using a LED source[24] and cameras using a uniform lump/LED illumination[25]. This offers an opportunity to evaluate the photon-number-resolving ability of a PMT. By THG 'imaging' of a coverslip interface (homogeneous sample) in a manner like the high-zoom 2PF/3PF 'imaging' of a fluorophore solution

(Fig. 1), we resolved >28 simultaneously arriving THG photons within a PDR upper limit of 28 effective photons (Fig. 2c), even though the excess noise exceeded 0.5 photoelectron in the *continuous* histograms of PMT output[23] (Fig 2a, Inserts). Alternatively (and at a higher cost), *discrete* photon-number resolving ability has been demonstrated in a point-like superconducting transition-edge sensor[26] and a quanta image sensor[27] by lowering the readout noise of photo-detection below 0.5 photoelectron.

*Multiphoton illumination fields visualized in bulk solutions*
Using the same eSLAM or pSLAM microscope, we switched from the above *in situ* absolute measurement to the visualization of illumination field[3] in fluorophore solutions, by simply switching the high-zoom scanning of <10×10 μm$^2$ to a low-zoom imaging across ~250×250 μm$^2$ with a frame size of 1024×1024 pixels (Fig. 1). Not surprisingly, the NADH solutions showed microscope-dependent but $C$-independent illumination fields pixelated with offset-removed arbitrary intensity values[3], indicating the off-axis effects such as higher-order field curvature (Fig. 3a,b) as opposed to on-axis illumination (Fig. 1, bottom left). Within the same microscope (eSLAM), comparative results using NADH and FAD solutions also revealed the dependence of illumination field on $n$ and modality/color/channel (Fig. 3b,c). However, varying-$P$ (varying-$C$) experiments using NADH solution(s) revealed no dependence of the illumination field on $P$ or signal strength ($C$ and imaging depth variation), indicating the reliability to visualize the illumination field in a bulk solution (Supplementary Fig. 7). In fact, our custom-built microscopes had relied on the 'flattening' of observed illumination field to optimize the alignment of relaying optics between laser source and photo-detection module.

The optimized 3PF/NADH field remained rather uneven due to high photon-order illumination, even though its predicted linear (one-photon) field was acceptably flat (Fig. 3d). This unevenness might have limited the reported FOV (123×123 μm$^2$) of deep 3PF imaging[30]. Also, the comparison between the observed 2PF/FAD field with the predicted two-photon field of the observed 3PF/NADH field highlighted the noticeable dependence of the field illumination (Fig. 3d, blue curves) on modality-specific detection path and efficiency[31] (Fig. 1). This complexity from multi-color detection, along with the non-absolute measurement of the illumination fields (with sample-dependent PMT gains and other device settings of digitizer, amplifier, and/or attenuator), have complicated the isolated quantity control of flat-field illumination using fluorescent slides[32] and bulk solutions[33].

*Pixelating with concentration-encoded effective photons (PCEP)*
Three key quality-control subtasks for reproducible and quantitative light microscopy are Poisson noise-limited detection, stable illumination, and flat-field illumination, which have been treated as independent goals[5,6] (Supplementary Table 1). After performing the three subtasks in two procedures detailed above, we attempted to integrate them into one simple procedure. Some previous attempts employed inhomogeneous samples and thus disengaged the flat-field illumination[22,23]. To engage this subtask, we quantitatively analyzed the pixelated arbitrary intensity values from the varying-$C$ calibration (Fig. 3b) and revealed the dependence of PDR on the size and location of region-of-interest (ROI) inside the FOV (Supplementary Fig. 8).

We found an optimal size of 30 pixel × 30 pixel or 7.5 μm × 7.5 μm, termed as a super-pixel, which was small to ensure a uniform ROI (required for Eq. 1) but large to generate a statistically convergent

PDR across the FOV (Fig. 3b, right; Supplementary Fig. 8). Because the $f_c$ parameters associated with different super-pixels scale with locally averaged illumination field strengths, the information of one constant $C$ is 'encoded' as different effective photons for individual pixels (Fig. 3b; Fig. 3d vs. Fig. 3e) after converting the arbitrary intensity values to effective photons (Supplementary Figs. 4 and 5, middle panels). In other words, the same effective photons at different pixels most likely represent different local $C$ in a bio-specimen, depending on their locations in the uneven illumination field. We thus term this integration 'pixelating with concentration-encoded effective photons' (PCEP), which departs greatly from previous *in situ* measurements of homogeneous samples[20,24].

Unexpectedly, the lower limit of the PDR associated with an on-axis super-pixel (~0.01 effective photon), i.e., detection limit due to the onset of specific non-Poisson noise that cannot be lowered by an increased $m$ (Fig. 2b), can be lowered by an increased off-axis extent down to ~0.001 effective photon (Fig. 3b, arrowheads). This effect suggests that the non-Poisson noise originates from local illumination field (which is worth future more detailed studies) rather than a constant additive noise[23] dictated by electronic settings such as the temporal window. Our PCEP produced an upper limit of ~0.001 effective photon for the additive noise. The low illumination field strength (i.e., low signal) of off-axis super-pixels in comparison to on-axis super-pixels is countered by an enhanced photo-detection (absence of non-Poisson noise) to homogenize the detection dynamic range in $C$ (~0.1-20 mM NADH) across all super-pixels (Fig. 3b, right). The resulting detection limit of ~0.1 mM NADH is termed as non-Poisson noise-equivalent concentration (NPNEC).

Thus, a misalignment-induced off-axis field illumination may produce a seemingly low detection limit in effective photons by the *in-situ* measurement of analog photo-detection, which would be mistakenly attributed to a high detection performance. This observation necessitates our integration of the three subtasks by PCEP. Also, the observed off-axis-enhanced photo-detection is especially beneficial for quantitative imaging of weak signals, using a strategy that rapidly acquires and averages single low SNR frames with either low-$P$ illumination or low-$C$ fluorescence labeling. This strategy mitigates the need to shrink the FOV[30] despite the highly uneven 3PF field illumination (Fig. 3d). A similar strategy has been appreciated for reduced phototoxicity but not for quantitative imaging[34].

*Quality control and performance benchmarking*
From the perspective of routine quality control, a varying-$P$ calibration (Fig. 2c) is simpler than the varying-$C$ calibration to perform PCEP, because it retains the absolute measurement of the latter using one constant $C$. After a varying-$C$ calibration in eSLAM (Fig. 3b) and 2 months of frequent biological imaging, we conducted one varying-$P$ calibration that confirmed its equivalence to the varying-$C$ calibration (Fig. 3f), except for the absence of lower detection limit that required a more accurate power-meter. The $f_P$ parameters associated with various super-pixels scaled with the corresponding $f_C$ parameters, while the continuous illumination field in effective photons approximated its varying-$C$ counterpart (Fig. 3e). It is thus feasible to ensure reproducible 3PF/NADH imaging via the varying-$P$ calibration at different time points, by first obtaining the same illumination field pixelated with effective photons and then the upper limit of additive noise from the weakest super-pixel. The former ensures no drift in optical alignment while the latter is necessary to monitor the aging of a point-like photodetector without interference from field illumination.

To extend the single-color PCEP of 3PF/NADH to the multi-color/modality detection of eSLAM with different PMTs (Supplementary Table 3), we performed additional varying-$P$ calibrations for 2PF/FAD, SHG, and THG imaging (Supplementary Fig. 5). We thus established a photon crosstalk matrix to quantify the color bleed-through among 4 modalities regardless of PMT gain and other detection settings (Fig. 1, see Methods). The flexibility in PMT gain allows tunable detection sensitivity and dynamic range (Supplementary Fig. 9) for different biological samples or applications but would prevent an arbitrary intensity analogue of this crosstalk matrix for objective quantification. In fact, the photon crosstalk matrix had guided our selections of excitation bands, dichroic mirrors, and optical filters (Fig. 1) in a feed-back process to minimize signal crosstalk while retaining detection efficiency.

We measured the point spread function (PSF) of eSLAM using ~100-nm fluorescent beads[35] and confirmed near diffraction-limited lateral-axial resolution (Supplementary Fig. 10). Because any degradation of PSF will weaken the effective photon-pixelated illumination field, this PSF measurement becomes a dependent subtask that only requires one-time quality control effort, which can be guaranteed if no change is detected from routine PCEP calibrations. Thus, an automatic 3D microscope stage with repeatable positioning[5] (required for the PSF measurement) becomes an optional quality-control subtask, as it is neither needed in PCEP (using a manual 1D stage) nor in some applications free of multi-FOV stitching (e.g., cell culture-based drug testing and clinical imaging[8,9]). In this way, 6 major quality-control subtasks[6] can be reduced to one routine procedure (Supplementary Fig. 1) especially beneficial for portable imaging.

To test the benchmarking by PCEP, we compared the $C$-encoded effective photons from PCEP and those from time-tagged computational photon-counting[28,29] in the experiment of Fig. 2b. The two independent methods yielded consistent results except for a proportional factor of 1.46 (Fig. 3g), which requires more detailed studies to understand. Overall, eSLAM attained an upgrade over SLAM to perform fluorescence lifetime imaging microscopy (FLIM)[28,29] (Supplementary Fig. 11). Using safe illumination powers that empirically avoided the phototoxicity of elevated auto-fluorescence[14,16], we compared the performance of SLAM-based and conventional multiphoton microscopes in optical metabolic imaging of NADH and FAD[36]. We identified a surprising advantage of eSLAM, i.e., a higher NADH imaging sensitivity by 3PF over 2PF (Supplementary Table 3). It is this pixelwise ability to encode $C$ that separates our study from a reported performance comparison among optical sectioning microscopes, which also converted the arbitrary intensity values of point- and camera-like detectors to effective photons[37]. To test the applicability of PCEP to camera-like detectors, we performed the varying-$C$ NADH calibration on EMCCD camera of a total internal reflection fluorescence (TIRF) microscope (Supplementary Fig. 7). The validity of PCEP was confirmed with an NPNEC of ~0.5 mM under the assumed safe illumination (Fig. 3h, arrowhead).

*Biological demonstration via eSLAM imaging*
To demonstrate the quantification by PCEP, we conducted eSLAM imaging on unlabeled cells and extracellular components of an *ex vivo* rabbit kidney (Fig. 4a). By performing arbitrary-intensity-value -to-effective-photon conversion established by PCEP and ignoring the difference of modality-dependent illumination fields (Fig. 3d, blue curves) as a first approximation, we employed the calibrated photon crosstalk matrix (Fig. 1) to correct across-modality color bleed-though and quantified the kidney cells across the modalities of 2PF/FAD, 3PF/NADH, and THG (Fig. 4b). The weak SHG signal from basement membrane-like collagen, which would otherwise be obscured by a strong 2PF bleed-through, became clearly discernible (compare Fig. 4a and Fig. 4b, green contrast). Importantly, gradual photo-bleaching occurred in 2PF and 3PF (full FOV) without the phototoxicity of elevated auto-fluorescence[16] during time-lapse imaging (Fig. 4c). The frame acquisition time (~0.33 s) of eSLAM using a resonant-galvanometer scanner allowed real-time monitoring of photo-bleaching and phototoxicity, in contrast to SLAM with a slow galvanometer-galvanometer scanner.

We produced the corresponding FLIM images and phasor plots for 2PF/FAD and 3PF/NADH signals (Fig. 4d). Interestingly, two different kidney tubules distinguishable by 2PF intensity and lifetime can be attributed to the proximal and distal tubules[38] with presumably different cellular metabolism (Fig. 4b,d, arrowheads). The FLIM data not only distinguishes the two tubules better than 2PF intensity (Fig. 4d, star), but also uniquely identifies the hemoglobin Soret fluorescence (peaked at 438 nm) from blood cells by its ultrashort lifetime[39] (Fig. 4d, arrows). Local absolute concentrations of FAD (or NADH) can be derived from the pixelated $C$-encoded effective photons via the phasor plots[40] (Supplementary Note 1) and corrected for the uneven 2PF/3PF field (Fig. 4e, left). The corresponding image of optical redox ratio $C_{FAD}/(C_{NADH}+C_{FAD})$ is thus obtained with rather small dependence on the field correction (Fig. 4e, right panels). Further refinement of this metabolic imaging is needed to discriminate NADH against NADPH (or FAD against cellular lipofuscin) in the 3PF (or 2PF) modality[36].

To reveal the enabling role of PCEP in microscopy-biology interaction, we tested eSLAM in numerous cell/tissue specimens and validated the safe illumination power (Supplementary Table 3) while confirmed no saturation (Fig. 2a, Insets) under a moderate PMT gain. We obtained the largest biological signal from mouse skull (THG up to 11 effective photons per pulse), which would saturate the photon-counting in SLAM under the same excitation (Supplementary Fig. 12). Also, the generally low cellular NADH signal in comparison to cellular FAD signal favors 1110 nm (eSLAM) over 1030 nm (pSLAM) for excitation (Supplementary Table 3). This performance benching creates the need to build a portable eSLAM microscope. Moreover, the observed dependence of illumination field on free-space detection path/modality (Fig. 3d, blue curves) points to an attractive alternative of optical fiber-coupled 16-channel spectral detection module[41] free of this dependence (Fig. 1, upper right), which would also permit tunable excitation[19]. In this prototypical process to optimize nonlinear optical imaging, PCEP allowed a custom-built microscope to co-evolve with the biology of interest toward reproducible, quantitative, gentle, and portable imaging.

**Discussion**

The quality-control tool of PCEP is generally applicable to molecular optical sectioning microscopy with a well-defined planar illumination field. With PCEP, image processing can be limited to rather simple tasks such as shading correction[31] (Supplementary Fig. 13), without any deconvolution or reconstruction that may prevent real-time visualization and/or quantitative analysis. Although we have only demonstrated PCEP in laser-scanning multiphoton microscopy and wide-field TIRF microscopy, the underlining mechanism may be broadly applicable to confocal microscopy[42] and light-sheet microscopy[33]. This is timely due to the recent standardization of laser-scanning confocal microscopy[43]. PCEP requires only a bulk fluorophore solution as standard sample, and thus avoids the special preparation of thin (~200 nm) and flat uniform fluorescent samples[31]. After placing the illumination field inside the solution just like how biological imaging is done, the varying-*P* PCEP calibration becomes a simple procedure widely supported by commercial microscopes. The procedure may be automated for routine self-diagnostic quality control of a core (static) imaging facility. This automation will be particularly beneficial for a portable imaging facility[7-9] because the embedded absolute measurement allows sensitive detection of any changes to laser source, photo-detection module, and relaying microscope optics.

Routine PCEP calibrations enable image pixel representation by the effective photons within a measured PDR (rather than an arbitrary intensity value). The error at each pixel is thus the squared root of the effective photons. This not only enables objective assessment of image quality but also supports a standard image format (Supplementary Table 1). Storing images in a standardized format[44] permits quantitative comparison of images from not only the same microscope over time, but also diverse microscopes with rich sample types, molecular targets, and imaging contrasts (e.g. fluorescence, harmonics[12], and molecular vibration[19]), allowing buildup of image archives for large-scale reanalysis[1,2]. Also, PCEP may empower photon-counting detection[12,23,24] to measure PDR and encode *C*, so that a specific imaging experiment can be reproduced by obtaining the same effective photon-pixelated image regardless of photo-detection mode. The resulting *C*-encoded effective photons may be correlated with the local concentration of an intrinsic[40] or fluorescence-labeled biomolecule of interest[3]. Beyond single-color imaging, proper PCEP calibration can quantitatively correct the color bleed-through in multicolor imaging of multiple biomolecules. Finally, by taking account of phototoxicity, PCEP can benchmark the performance of different modalities or microscopes to image the same molecular targets.

Beyond the quality control of a preexisting microscope, PCEP can serve as a precision measurement tool to optimize a custom-built microscope, with a detection limit down to 0.001 effective photon and a dynamic range of more than 3 orders of magnitude for a typical PMT at one gain (Supplementary Fig. 14). Our PCEP-assisted upgrade of SLAM (photon-counting) to eSLAM (analog photo-detection) transforms a low signal rate of less than one effective photon per pulse to a high signal rate of up to 11 simultaneously arrived effective photons per pulse under safe biological illumination. The 'idealized' design of eSLAM may serve as the starting point to generalize PCEP to diverse designs of molecular optical sectioning microscopy. In this process, PCEP may play a similar enabling role to optimize key elements of microscopy (choice of imaging modality, illumination *P*/field/photon-order, photo-detection mode/gain/color, microscope objective, zoom or FOV, pixels per frame, frame acquisition time, etc.), by linking them with interacting biological elements (fluorophore *C*, photo-bleaching, targeted or labeled biomolecules, samples of interest, intended spatial/temporal resolution, phototoxicity, etc.). At large imaging depths, PCEP-calibrated quantification becomes susceptible to the interference of out-of-focus background in deep imaging[30], which depends on sample absorption-scattering properties and thus requires specific protocol to estimate[37]. One way to navigate this challenge is to work with thin sectioned samples[45] and thin homogeneous fluorescent phantoms[31], so that PCEP-based quality control may be extended to the wide-field epi-fluorescence microscopy with non-laser light sources.

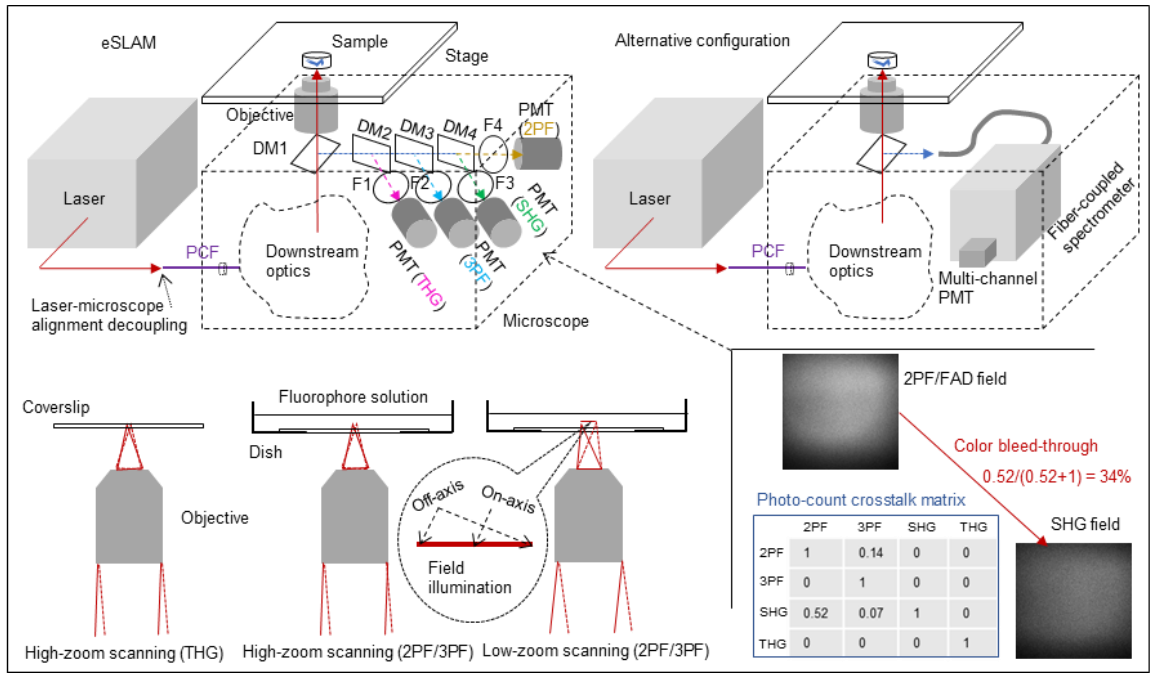

**Fig. 1: Schematic of eSLAM with built-in quality control**. The inverted microscope (see Supplementary Fig. 2 for more details) consists of a source femtosecond laser, a spectrum-broadening module based on photonic crystal fiber (PCF), subsequent relaying optics with a mechanical stage to perform high/low-zoom 2PF/3PF 'imaging' of a fluorophore solution at 15±5 μm depth and THG imaging of coverslip interface (bottom left), and photo-detection paths with specific dichroic mirrors (DM) and optical filters (F) corresponding to 4 modalities (THG, 3PF/NADH, SHG, and 2PF/FAD), along with an alternative configuration with optical fiber-coupled spectral detection module (upper right). The optical alignment of the laser and the microscope is decoupled (i.e., laser-microscope alignment decoupling) because the misalignment of the former can be easily detected by the altered output spectrum of the PCF (see ref. 19 for details). Inset: measured bleed-trough from illumination field of 2PF/FAD to that of SHG and resulting photon crosstalk matrix for 4 modalities.

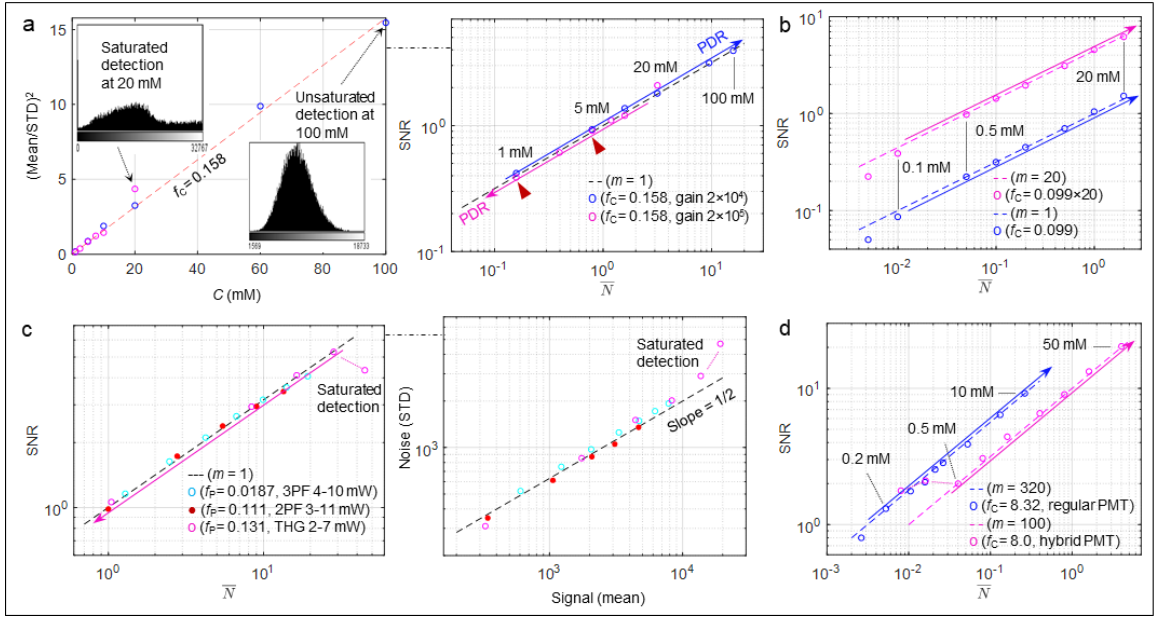

**Fig. 2:** *In situ* absolute measurement of PDRs using stable illumination. **a**, Parameter $f_C$ obtained from single-parameter linear fit (broken line) between experimental pSLAM-3PF (mean/STD)$^2$ and $C$ (NADH concentration) in two varying-$C$ ($P$ = 2.4 mW) experiments at two different PMT gains (colored points), with unsaturated/saturated histogram at low/high gain (Insets); right panel shows the corresponding log-scale PDRs from Eq. 2 (colored points and arrowed lines), in which both PMT gains attain 0.16 and 0.79 effective photon at 1 mM and 5 mM, respectively, along with prediction from Eq. 1 (broken line). **b**, PDRs from varying-$C$ (NADH) constant-$P$ (16.8 mW) eSLAM (Eq. 2) at PMT gain 4.8×10$^5$ corresponding to one frame and 20 frames (colored points and arrowed lines) along with prediction from Eq. 1 (broken lines). **c**, PDRs from varying-$P$ pSLAM (Eq. 3) at PMT gain 2×10$^4$ corresponding to 10 mM NADH/3PF, 10 mM FAD/2PF, and coverslip/THG (colored points and arrowed line) along with prediction from Eq. 1 (broken line); right panel shows their counterparts of photon transfer curves (PTCs). **d**, PDRs from varying-$C$ (NADH) constant-$P$ (15 mW for regular PMT, 30 mW for hybrid PMT) traditional multiphoton microscopy (Eq. 2) corresponding to a regular PMT (H7422P-40, Hamamatsu) at gain 1.1×10$^6$ and a hybrid PMT (R10467U-40, Hamamatsu) at gain 1.2×10$^5$ (colored points and arrowed lines) along with prediction from Eq. 1 (broken lines).

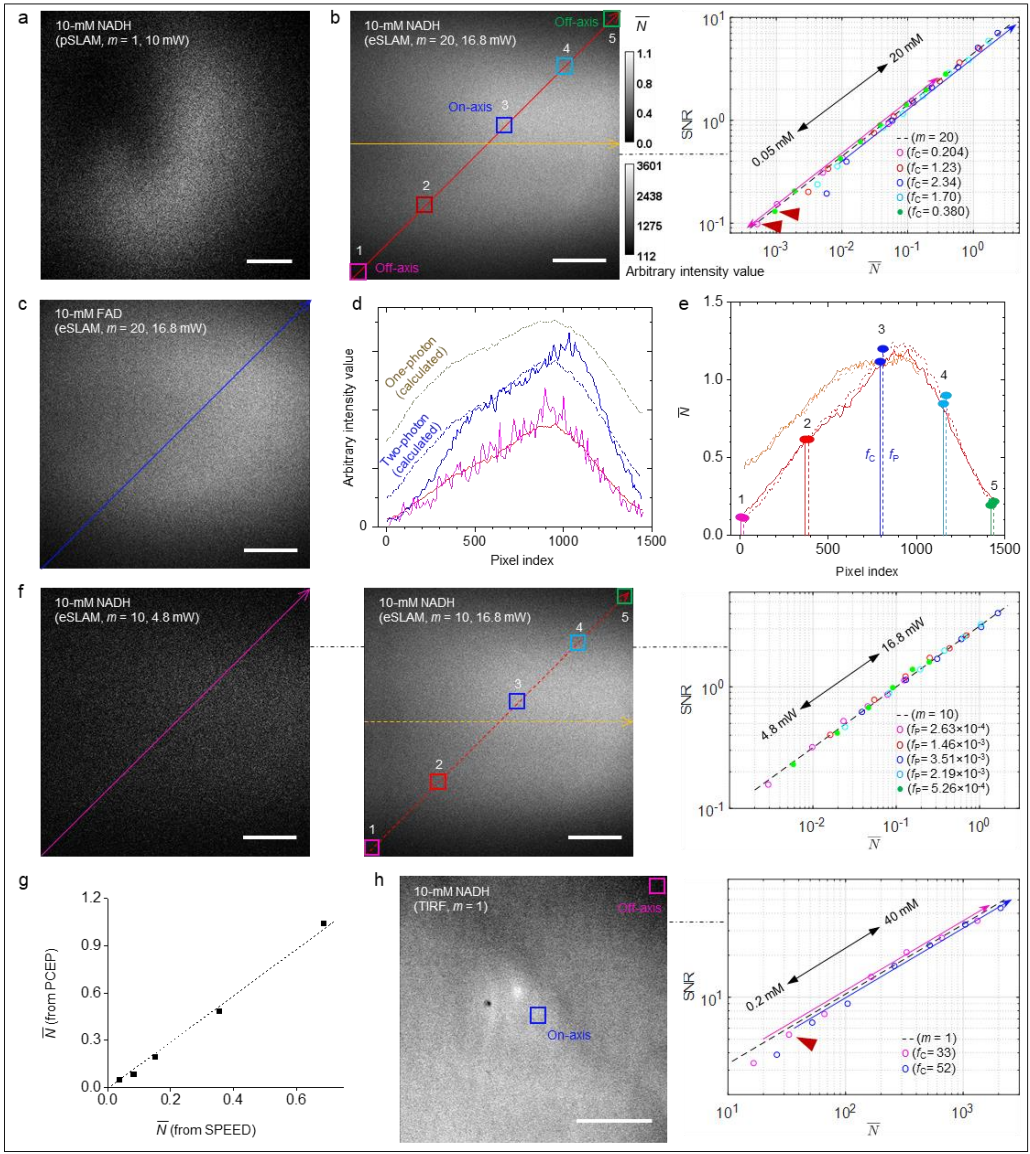

**Fig. 3: Characteristic features of PCEP. a**, pSALM illumination field (3PF) in NADH solution. **b**, eSLAM illumination field (3PF) visualized in NADH solution with predefined lines and super-pixels (boxes); right panel shows PDRs from varying-$C$ (NADH) constant-$P$ experiment (Eq. 2) at PMT gain $4.8\times10^5$ corresponding to different super-pixels (colored points and arrowed lines) and prediction from Eq. 1 (broken line). **c**, eSLAM illumination field (2PF) visualized in FAD solution. **d**, Diagonal line profiles in **b**, **c**, **f**-left (solid lines) and calculated 1- or 2-photon counterpart of 3PF/NADH profile in **b** (broken lines). **e**, Diagonal/lateral line profiles in **b**-left and **f**-middle consistent with properly scaled $f_C/f_P$ parameters (vertical segments). **f**, eSLAM illumination field visualized in NADH solution using low $P$ (left) or high $P$ (middle); right panel shows SNR vs. $\bar{N}$ relation from varying-$P$ experiment (Eq. 3) at PMT gain $4.8\times10^5$ corresponding to different super pixels (colored points) and prediction from Eq. 1 (broken line). **g**, Comparison of effective photons in a vary-$C$ NADH experiment obtained from either PCEP or single/multi-photon peak event detection (SPEED). **h**, TIRF illumination field (one-photon fluorescence) visualized in NADH solution at 405-nm excitation with on-axis and off-axis super-pixels; right panel shows PDRs from varying-$C$ (NADH) constant-$P$ experiment (Eq. 2) corresponding to the two super-pixels (colored points and solid lines) and prediction from Eq. 1 (broken line). Scale bars: 50 μm.

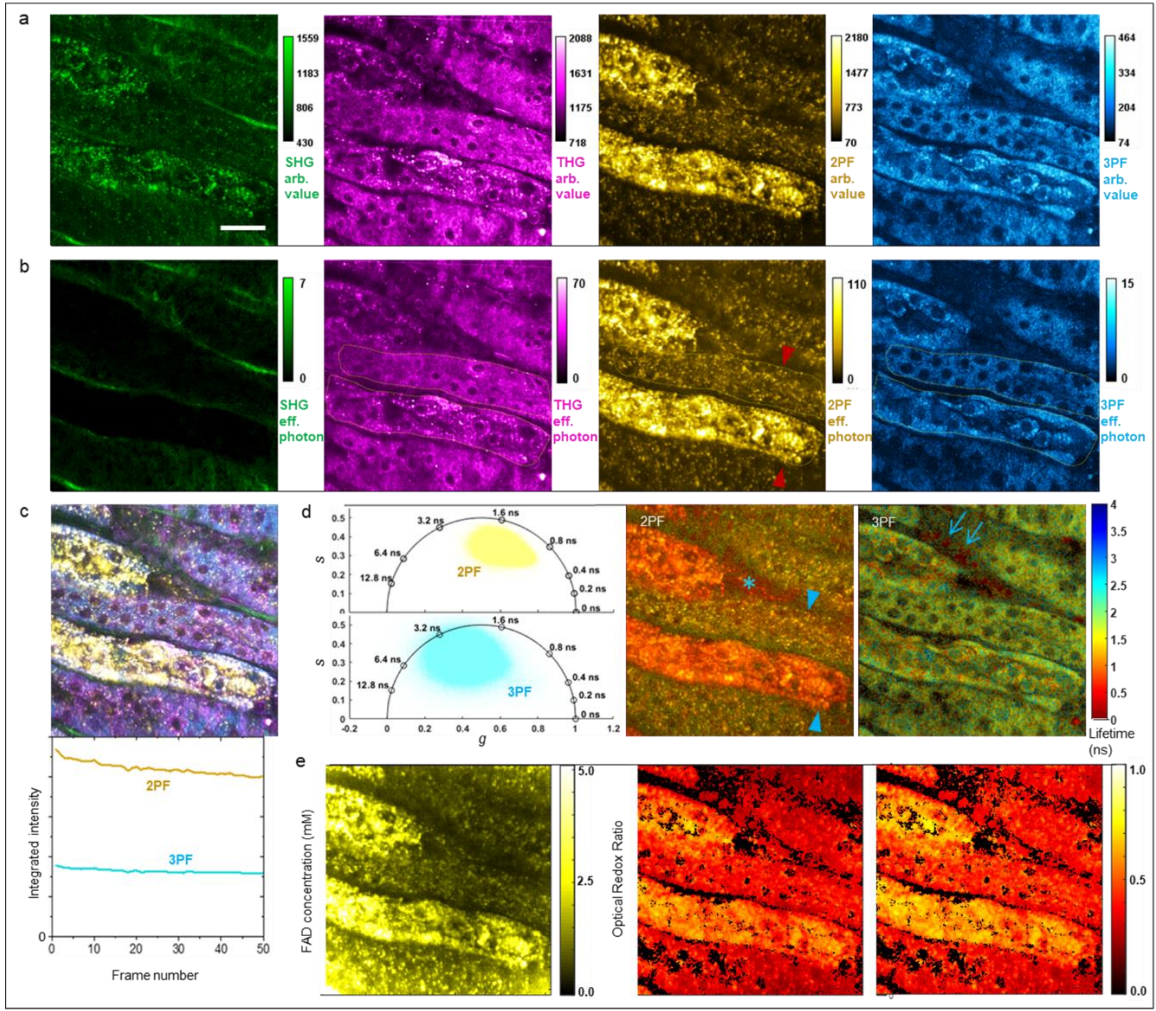

**Fig. 4: Quantitative biological imaging by eSLAM.** Scale bar: 50 µm. **a**, Arbitrary intensity value-pixelated images of *ex vivo* rabbit kidney at ~15-µm imaging depth with SHG, THG, 2PF/FAD, and 3PF/NADH signals (50-frame summation). **b**, Corresponding effective photon-pixelated images after color bleed-through correction showing kidney epithelial cells with average effective photons per pixel of 27 (THG), 30 (FAD), and 3 (NADH). **c**, Composite image (top) and real-time monitoring of FAD/NADH photo-bleaching during time-lapse imaging (bottom). **d**, phase plots (left) and corresponding FLIM images of FAD and NADH (right) showing fluorescence lifetime with intensity overlay (9-pixel averaging over neighboring pixels). **e**, Image of FAD concentration corrected for uneven field illumination (left) and related images of optical redox ratio with (middle) and without the field correction (right).

## Methods

### Optical setup of eSLAM

Details of the laser source have been reported elsewhere[15]. The 5 MHz supercontinuum pulses from this source were sent into a 128-pixel 4f pulse shaper (femtoJock Box, BioPhotonic Solutions Inc.) to select an excitation band of 1110±30 nm. The spectrally selected pulses were then raster scanned by a resonant mirror (10×10 mm, 1592 Hz line rate, EOCP) and a galvanometer mirror (GVS011, Thorlabs), and finally focused by a microscope objective (UAPON 40XW340, N.A. = 1.15, Olympus) with up to ~35 mW average power on the sample. A pair of identical achromatic doublets (AC254-050-C-ML – f=50 mm, Thorlabs) and another pair of different achromatic doublets (AC254-030-C-ML – f=30 mm, AC508-100-C-ML – f=100 mm, Thorlabs) were used for 4f telecentric resonant-galvanometer beam steering, while the latter also expanded the beam to fill the back focal pupil plane of the objective (Ø10.35mm). The actual/safe power on the sample was adjusted by a neutral density (ND) filter while the corresponding pulse width was compressed to near-transform-limited value (~60 fs, FWHM) by the pulse shaper[19]. Average incident power at the sample plane was measured by a microscope slide power meter (S175C, Thorlabs). The photo-detection of eSLAM followed that of SLAM except for the replacement of photon-counting PMTs with analog PMTs (Supplementary Table 3). The whole system functioned as an inverted multiphoton microscope.

### Signal acquisition and processing (general)

The pulse repetition rate of the laser source (40 MHz) was divided to 10 MHz by a frequency divider (PRL-260BNT, Pulse Resaerch Lab) and distributed by a fanout line driver (PRL-414B, Pulse Resaerch Lab), and then used as the master clock to synchronize the resonant mirror and subsequent signal acquisition. For the resonant mirror, the active acquisition length was designed to occupy the central 65% of the sinusoidal line profile (spatial fill fraction), with one pulse per pixel per frame (i.e. pixel dwell time 0.2 μs). The PMT-detected 2PF and 3PF signals were first sent to high-speed current-to-voltage conversion amplifier unit (C5594-12, Hamamatsu) with 1.5 GHz cutoff frequency. The converted voltage signals were then digitized by a 2 GHz dual-channel high-speed digitizer (ATS9373, AlazarTech). For high dynamic range calibration of photo-detection using a NADH/FAD solution, a 20dB attenuator was connected after the amplifier to match the range of digitizer input voltage (±400 mV). The signals from SHG and THG modalities were amplified by a 60 MHz bandwidth amplifier (TIA60, Thorlabs) and digitized by a 125 MS/s digitizer (ATS9440, AlazarTech).

A GPU (GeForce RTX 2080, NVIDIA) enabled real-time image display and accelerated raw data process. Our design supported a maximal frame (1024 pixel × 1024 pixel) rate of 3 Hz by bidirectional resonant scanning but was limited to ~1.7 Hz by the storage of rapidly digitized 2PF and 3PF modalities. At 5 MHz repetition rate and 2 GS/s sampling rate, there are 400 sampling points between pulses. Because the fluorescence lifetime of FAD or NADH is less than 10 ns (20 sampling points), at least 95% of the data points are noise points. To avoid these noise points, the position of the maximum value within each line of fast scan was first found by superimposing the raw data of all the pixels in the line. Then, the custom data points near the maximum value position (time-gated window) were extracted using a custom LabVIEW-based GUI with 40 data points per pulse, 9 before the maximum value and 30 after the maximum value (Supplementary Fig. 15). By implementing this algorithm in the GPU, most noise points were removed before storage.

Before performing PCEP, the offset values must be removed from all pixels of field illumination images[3]. For the vary-$C$ (varying-$P$) experiment, the offset values are obtained from a field illumination image of blank control solution of $C = 0$ (or a fluorophore solution at $P = 0$). Then, parameter $f_C$ (or $f_P$) is determined by the single-parameter linear fit between experimental (Mean/STD)$^2$ from a small FOV or a ROI (or super-pixel) of field illumination images and $C$ (or $P^n$). This produces an experimental SNR versus $\bar{N}$ relation (Eq. 2 or Eq. 3) to compare with the theoretical relation with known $m$ (Eq. 1), resulting in a measured PDR often with a distinguishable lower end that exhibits disagreement.

*Signal acquisition and processing (FLIM)*

To validate the FLIM capability of eSLAM, the lifetimes of SHG (~0.0 ns), NADH in 1M HEPES (~0.4 ns), Rhodamine B in water (~1.7 ns), and Fluorescein in ethanol (~3.4 ns) were tested using computational photon counting by employing the single- and multi-photon peak event detection (SPEED) algorithm[28,29]. Specifically, Rhodamine B and Fluorescein were used to calibrate the 2PF modality, NADH was used to calibrate the 3PF modality, and the temporal impulse response function (IRF) of the system was determined by SHG imaging of a urea crystal (Supplementary Fig. 11, top). The estimated fluorescence lifetimes approximated the known values (Supplementary Fig. 11, bottom). However, low fluorescence lifetimes (such as NADH and SHG) were biased slightly higher due to the limited bandwidth of collection electronics, and longer fluorescence lifetimes (Rhodamine B and Fluorescein) were biased to lower values, likely due to the low probability of collecting and properly time-tagging later-arriving photons when using one laser pulse per pixel and inferring the laser pulse synchronization[29]. The IRF of photo-detection had a full-width-half-maximum (FWHM) of 0.56 ns, slightly higher than previously reported IRFs estimated using SPEED which used a different PMT and faster digitization rates (0.32 ns IRF with 3.2 GS/s ADC[28], 0.23 ns IRF with 5 GS/s ADC[29]). Despite these biases, existing eSLAM sufficiently revealed relative changes in fluorescence lifetime over the range of interest for NADH and FAD (Fig. 4d).

Both SPEED and PCEP were used to quantify the average number of photons per pulse using different concentrations of NADH from 0.5 to 20 mM. Data was analyzed to determine the average number of photons per pulse per 1 mM NADH, resulting in 0.060 photons/pulse/1 mM NADH for SPEED and 0.068 photons/pulse/1 mM NADH for PCEP with a strong linear correlation ($R^2 = 0.998$, Fig. 3g). A small percentage (<10%) of photon counts are being missed at the higher photon rates of the experimental data using SPEED, leading to values that are biased slightly lower due to the finite dead time of the system[28,29] (~1.0 ns, or twice the sampling rate). The SPEED method requires high sampling rate (>1 GS/s with high consumption on acquisition and computation) to restore the original signal and time tag detected photons for FLIM imaging, whereas PCEP simply uses the integrated/averaged signal collected at lower digitization rates and supports up to 23 effective photons per pulse (Fig. 2c) but does not time-tag the detected photons. The presented eSLAM system is capable of both SPEED and PCEP for quantification due to its versatile design.

*Photon crosstalk matrix*

The varying-*P* experiments for 4 different modalities (2PF, 3PF, SHG, and THG) not only calibrated the relation between pixel arbitrary intensity values and effective photons within individual modalities (Supplementary Fig. 5), but also produced the color bleed-through across these modalities due to the simultaneous signal acquisition by eSLAM. The input signal of a given modality and related bleed-through to other modalities (columned percentages in Fig. 1, Inset) were qualified using the converted effective photons from the arbitrary intensity values (Supplementary Fig. 5, middle panels) over the full field-of-view of illumination (Fig. 1, images). This removed any dependence of the resulting crosstalk matrix (Fig. 1, Inset) on illumination power, PMT gain, or other device settings (which would be present if the arbitrary intensity value were used). Although the calibration was performed at specific PMT gains, the arbitrary intensity values at different gains can be calculated by the ratio of actual versus calibrated gain to produce a look-up table of arbitrary-intensity-value -to-effective-photon conversion. Despite this complexity, the photon crosstalk matrix remains constant for different PMT gains (as demonstrated in Fig. 2a). It can be varied by changing the optical filters and dichroic mirrors of photo-detection module to minimize signal crosstalk (Fig. 1). The observed small crosstalk between 3PF and 2PF modalities (Fig. 1, Inset), not expected from the spectrally overlapped emission of NADH and FAD (Supplementary Fig. 2), is due to clipping-assisted dual-fluorophore sensing[46].

The photon crosstalk matrix in Fig. 1 was then divided by the total applied load in each modality to obtain the transfer function *K*:

$$K = \begin{vmatrix} 0.657 & 0.112 & 0 & 0 \\ 0 & 0.828 & 0 & 0 \\ 0.343 & 0.060 & 1 & 0 \\ 0 & 0 & 0 & 1 \end{vmatrix}$$

and the inverse matrix $K^{-1}$:

$$K^{-1} = \begin{vmatrix} 1.523 & -0.206 & 0 & 0 \\ 0 & 1.207 & 0 & 0 \\ -0.523 & -0.0012 & 1 & 0 \\ 0 & 0 & 0 & 1 \end{vmatrix}$$

Applying $K^{-1}$ to each pixel of eSLAM images (Fig. 4a) after arbitrary-intensity-value-to-effective-photon conversion resulted in the crosstalk-compensated images without field correction (Fig. 4b).

### *Overview of calibration and data processing in eSLAM*

Software components of eSLAM have been discussed individually. With PCEP-based calibration that determine various data processing parameters, the digitized analog outputs of PMTs from a biological sample can be ultimately converted to effective photon-pixelated multimodal eSLAM images, and related fluorescence lifetime images and concentration images of targeted fluorophores (Supplementary Fig. 16). Thus, eSLAM can serve as a starting point of quantitative imaging for diverse designs of molecular optical sectioning microscopy (Supplementary Fig. 3).

### *Other microscopes and stage focusing*

Details of our SLAM microscope, pSLAM microscope, and traditional multiphoton microscope (MPM) have been reported in one paper[12], a preprint[14] and two recent reports[28,29], respectively. The main independent parameters of these laser-scanning microscopes are compared with those of eSLAM

(Supplementary Table 3). The wide-field inverted TIRF microscope built upon a Zeiss Axiovert 200M microscope is equipped with an oil immersion microscope objective (63x, NA 1.4) and a cooled Photometrics 512 Evolve EMCCD camera to image the fluorescence of single molecules or a thin (< 200 nm typically) specimen.

The stage focusing of these microscopes was required for PCEP to visualize the illumination field in a bulk fluorophore solution inside a 35-mm glass (coverslip) bottom dish (Fig. 1). For all laser-scanning inverted microscopes (eSLAM, pSLAM, and MPM), the interface between coverslip and the solution was detected by continuous low-zoom fluorescence imaging while manually positioning a microscope stage. Then, the illumination plane was placed ~10 μm inside the solution using the stage. For the TIRF microscope, this interface was first marked with fiducial lines by a diamond knife and detected by the built-in bright-field imaging of the same microscope. The bright-field imaging was then switched to TIRF imaging while the corresponding illumination field was placed ~10 μm inside the solution manually using the built-in stage of the microscope.

### *Preparation of bulk solutions*
NADH (Grade I, Sigma) and FAD (94% dry wt., Thermo Scientific) were dissolved in a 1 M HEPES buffer to maintain a stable pH. Rhodamine B and acridine orange were dissolved in sterile water. Fluorescein was dissolved in 100% ethanol.

### *Animal tissue*
All animal procedures were conducted in accordance with protocols approved by the Illinois Institutional Animal Care and Use Committee at the University of Illinois at Urbana-Champaign. The internal organs were obtained from ~3-month-old 2.8 kg laboratory female New Zealand white albino rabbits (Oryctolagus cuniculus) (Charles River Laboratories, Wilmington, MA) bearing subcutaneous rabbit mammary tumors within 10 minutes post-mortem. The excised kidneys (or hearts) were immediately submerged in sterile $Ca^{2+}/Mg^{2+}$-free 0.1 μm filter-sterilized PBS (pH 7.0 – 7.2) and washed from blood by changing the PBS solution 3 times. Each organ sample was manually sliced in axial and sagittal planes in sterile tissue culture dish kept on ice. Individual tissue slices were then placed onto the uncoated 35 mm imaging dishes with No. 0 coverslip and 20 mm glass diameter (MatTek, #P35G-0-20-C). The slices were incubated in 500 μL FluoroBrightTM DMEM (TFS, #A1896701) supplemented with 10% FBS, 1% PSA, and 4 mM L-Glutamine solutions. Mice (C57BL/6J, Jackson Laboratory) were used to obtain *ex vivo* skull samples, which were imaged directly without solution-based preparation.

### *Cell culture*
Human breast cancer cells MCF7 (ATCC HTB-22) were maintained in EMEM supplemented 10% FBS, 5 μg/mL insulin and 1% penicillin streptomycin antibiotic, and grown in an incubator at 37 °C with 5% $CO_2$. One day prior to imaging, cells were plated on poly-D-lysine coated 35 mm diameter glass-bottom imaging dishes (P35GC-0-10-C, MatTek) and incubated overnight in 2 mL of media to adhere.

**Data availability**

The data that support the findings in this study are available from the corresponding author upon reasonable request.

## References

1. Ellenberg, J. et al. A call for public archives for biological image data. *Nature Methods* **15**, 849-854 (2018).
2. Williams, E. et al. Image Data Resource: a bioimage data integration and publication platform. *Nature methods* **14**, 775-781 (2017).
3. Waters, J.C. Accuracy and precision in quantitative fluorescence microscopy. *Journal of Cell Biology* **185**, 1135-1148 (2009).
4. Montero Llopis, P. et al. Best practices and tools for reporting reproducible fluorescence microscopy methods. *Nature Methods* **18**, 1463-1476 (2021).
5. Boehm, U. et al. QUAREP-LiMi: a community endeavor to advance quality assessment and reproducibility in light microscopy. *Nature Methods* **18**, 1423-1426 (2021).
6. Faklaris, O. et al. Quality assessment in light microscopy for routine use through simple tools and robust metrics. *Journal of Cell Biology* **221**, e202107093 (2022).
7. Power, R.M. & Huisken, J. Putting advanced microscopy in the hands of biologists. *Nature Methods* **16**, 1069-1073 (2019).
8. Balu, M. et al. In vivo multiphoton NADH fluorescence reveals depth-dependent keratinocyte metabolism in human skin. *Biophysical Journal* **104**, 258-267 (2013).
9. Kantelhardt, S.R. et al. In vivo multiphoton tomography and fluorescence lifetime imaging of human brain tumor tissue. *Journal of Neuro-Oncology* **127**, 473-482 (2016).
10. Eisenstein, M. Smart solutions for automated imaging. *Nature Methods* **17**, 1075-1079 (2020).
11. Lemon, W.C. & McDole, K. Live-cell imaging in the era of too many microscopes. *Current Opinion in Cell Biology* **66**, 34-42 (2020).
12. You, S. et al. Intravital imaging by simultaneous label-free autofluorescence-multiharmonic microscopy. *Nature Communications* **9**, 2125 (2018).
13. Žurauskas, M., Durack, M., Tu, H. & Boppart, S.A. in Advanced Biomedical and Clinical Diagnostic and Surgical Guidance Systems XX, Vol. 11949 41-45 (SPIE, 2022).
14. Wang, G., Boppart, S.A. & Tu, H. Compact simultaneous label-free autofluorescence multi-harmonic (SLAM) microscopy for user-friendly photodamage-monitored imaging. *arXiv preprint arXiv:2210.13556* (2022).
15. Wang, G. & Tu, H. Fiber-optic nonlinear wavelength converter for adaptive femtosecond biophotonics. *arXiv preprint arXiv:2305.08266* (2023).
16. Galli, R. et al. Intrinsic indicator of photodamage during label-free multiphoton microscopy of cells and tissues. *PloS ONE* **9**, e110295 (2014).
17. König, K., So, P., Mantulin, W. & Gratton, E. Cellular response to near-infrared femtosecond laser pulses in two-photon microscopes. *Optics Letters* **22**, 135-136 (1997).
18. Tu, H. & Boppart, S.A. Coherent fiber supercontinuum for biophotonics. *Laser & Photonics Reviews* **7**, 628-645 (2013).

19. Tu, H. et al. Stain-free histopathology by programmable supercontinuum pulses. *Nature Photonics* **10**, 534-540 (2016).
20. Benninger, R.K., Ashby, W.J., Ring, E.A. & Piston, D.W. Single-photon-counting detector for increased sensitivity in two-photon laser scanning microscopy. *Optics Letters* **33**, 2895-2897 (2008).
21. Cheng, L.-C., Horton, N.G., Wang, K., Chen, S.-J. & Xu, C. Measurements of multiphoton action cross sections for multiphoton microscopy. *Biomedical Optics Express* **5**, 3427-3433 (2014).
22. Driscoll, J.D. et al. Photon counting, censor corrections, and lifetime imaging for improved detection in two-photon microscopy. *Journal of Neurophysiology* **105**, 3106-3113 (2011).
23. Modi, M.N., Daie, K., Turner, G.C. & Podgorski, K. Two-photon imaging with silicon photomultipliers. *Optics Express* **27**, 35830-35841 (2019).
24. Ching-Roa, V.D., Olson, E.M., Ibrahim, S.F., Torres, R. & Giacomelli, M.G. Ultrahigh-speed point scanning two-photon microscopy using high dynamic range silicon photomultipliers. *Scientific Reports* **11**, 1-12 (2021).
25. Li, L., Li, M., Zhang, Z. & Huang, Z.-L. Assessing low-light cameras with photon transfer curve method. *Journal of Innovative Optical Health Sciences* **9**, 1630008 (2016).
26. Lita, A.E., Miller, A.J. & Nam, S.W. Counting near-infrared single-photons with 95% efficiency. *Optics Express* **16**, 3032-3040 (2008).
27. Ma, J., Masoodian, S., Starkey, D.A. & Fossum, E.R. Photon-number-resolving megapixel image sensor at room temperature without avalanche gain. *Optica* **4**, 1474-1481 (2017).
28. Sorrells, J.E. et al. Single-photon peak event detection (SPEED): a computational method for fast photon counting in fluorescence lifetime imaging microscopy. *Optics Express* **29**, 37759-37775 (2021).
29. Sorrells, J.E. et al. Computational Photon Counting Using Multithreshold Peak Detection for Fast Fluorescence Lifetime Imaging Microscopy. *ACS Photonics* **9**, 2748-2755 (2022).
30. Horton, N.G. et al. In vivo three-photon microscopy of subcortical structures within an intact mouse brain. *Nature Photonics* **7**, 205-209 (2013).
31. Zwier, J., Van Rooij, G., Hofstraat, J. & Brakenhoff, G. Image calibration in fluorescence microscopy. *Journal of Microscopy* **216**, 15-24 (2004).
32. Brown, C.M., Reilly, A. & Cole, R.W. A quantitative measure of field illumination. *Journal of Biomolecular Techniques: JBT* **26**, 37 (2015).
33. Hobson, C.M. et al. Practical considerations for quantitative light sheet fluorescence microscopy. *Nature Methods*, 1-12 (2022).
34. Chen, X. et al. LOTOS-based two-photon calcium imaging of dendritic spines in vivo. *Nature Protocols* **7**, 1818-1829 (2012).
35. Cole, R.W., Jinadasa, T. & Brown, C.M. Measuring and interpreting point spread functions to determine confocal microscope resolution and ensure quality control. *Nature Protocols* **6**, 1929-1941 (2011).
36. Kolenc, O.I. & Quinn, K.P. Evaluating Cell Metabolism Through Autofluorescence Imaging of NAD(P)H and FAD. *Antioxidants and Redox Signaling*, ars.2017.7451 (2017).
37. Murray, J.M., Appleton, P.L., Swedlow, J.R. & Waters, J.C. Evaluating performance in three-

dimensional fluorescence microscopy. *Journal of Microscopy* (2007).
38. Dunn, K.W., Sandoval, R.M., Kelly, K.J., Dagher, P.C. & Molitoris, B.A. Functional studies of the kidney of living animals using multicolor two-photon microscopy. *American Journal of Physiology Cell Physiology* **283**, C905 (2002).
39. Zheng et al. Two-photon excited hemoglobin fluorescence. *Biomedical Optics Express* (2011).
40. Ning, M., Digman, M.A., Malacrida, L. & Gratton, E. Measurements of absolute concentrations of NADH in cells using the phasor FLIM method. *Biomedical Optics Express* **7**, 2441 (2016).
41. Wahl, M. et al. Photon arrival time tagging with many channels, sub-nanosecond deadtime, very high throughput, and fiber optic remote synchronization. *Review of Scientific Instruments* **91**, 013108 (2020).
42. Jonkman, J., Brown, C.M., Wright, G.D., Anderson, K.I. & North, A.J. Tutorial: guidance for quantitative confocal microscopy. *Nature Protocols* **15**, 1585-1611 (2020).
43. International Organization for Standardization. Microscopes — Confocal microscopes — Optical data of fluorescence confocal microscopes for biological imaging. *ISO Standard* No. 21073 (2019).
44. Moore, J. et al. OME-NGFF: a next-generation file format for expanding bioimaging data-access strategies. *Nature Methods* **18**, 1496-1498 (2021).
45. Rivenson, Y. et al. Virtual histological staining of unlabelled tissue-autofluorescence images via deep learning. *Nature Biomedical Engineering* **3**, 466-477 (2019).
46. Boppart, S.A., You, S., Li, L., Chen, J. & Tu, H. Simultaneous label-free autofluorescence-multiharmonic microscopy and beyond. *APL Photonics* **4**, 100901 (2019).

**Acknowledgements**

The authors thank Dianwen Zhang for helping with TIRF experiments. This work was supported by a grant from the National Institutes of Health, U.S. Department of Health and Human Services (R01 CA241618).

**Author contributions**

G.W. and H.T. conceived the idea. H.T. and S.A.B. obtained the funding and supervised the study. R.R.I., J.E.S., J.S., and Y.S. developed parts of the methodology. R.R.I., J.E.S., and G.W. developed data acquisition software. G.W., J.E.S., R.R.I., E.A., C.A.R, E.J.C., and J.P. performed experiments. G.W., H.T., J.E.S., and R.R.I. performed data analysis. H.T. and G.W. drafted the manuscript. H.T. and S.A.B. reviewed and edited the manuscript with inputs from all authors.

**Ethics declarations**

G.W. and H.T. are contacting the Office of Technology Management at the University of Illinois at Urbana-Champaign for commercial potential of the developed technology.

**Supplementary Note 1**

Concentration values of NADH and FAD were estimated using a previously developed method that uses the intensity and fluorescence lifetime phasor plot[40] along with calibrated measurements. Briefly, a known amount of synthetic DC light (L) was added to x-y-t images in order to shift the location of the phasor coordinates from their proper locations (Supplementary Fig. 17a,c) to shifted locations (Supplementary Fig. 17b,d). Adding this DC light causes a change in modulation (M) and phasor variables g and s dependent on the fluorescence intensity (F):

$$M_{shifted} = M_{original} \times \left(1 - \frac{L}{L+F}\right)$$

$$g_{shifted} = g_{original} \times \left(1 - \frac{L}{L+F}\right)$$

$$s_{shifted} = s_{original} \times \left(1 - \frac{L}{L+F}\right)$$

Using known lifetimes ($\tau_{free-ref}$, $\tau_{bound-ref}$), their corresponding phasor variables ($g_{free-ref}$, $g_{bound-ref}$, $s_{free-ref}$, $s_{bound-ref}$), and intensities ($F_{free-ref}$, $F_{bound-ref}$) of free and bound NADH and FAD at 1 mM, each pixel within the image was fit to a concentration of NADH (Supplementary Fig. 17e,f) and FAD (not shown). This was accomplished by iterative nonlinear least squares fitting of a single pixel's location on the phasor plot to a concentration-equivalent line connecting linear combinations of free and bound fluorophore at equal concentrations. Both g and s coordinates of the free and bound fluorophore were calculated for a given concentration (C) as follows:

$$g_{bound} = g_{bound-ref} \times \left(1 - \frac{L}{L + C \times F_{bound-ref}}\right)$$

$$g_{free} = g_{free-ref} \times \left(1 - \frac{L}{L + C \times F_{free-ref}}\right)$$

$$s_{bound} = s_{bound-ref} \times \left(1 - \frac{L}{L + C \times F_{bound-ref}}\right)$$

$$s_{free} = s_{free-ref} \times \left(1 - \frac{L}{L + C \times F_{free-ref}}\right)$$

Then, the distance between the g and s values for the given pixel and the concentration-equivalent line was minimized using iterative nonlinear least squares fitting by adjusting concentration only.

**Supplementary Table 1.** 'Idealized' design of eSLAM to classify quality control subtasks of live-cell fluorescence imaging and reduce 15 isolated subtasks to a minimum effort of 3 independent subtasks.

| Subtask (reference) | Classification (related figure) | Reason for designated classification |
|---|---|---|
| Reproducible sample preparation[3,4] | Irrelevant | No sample preparation for label-free imaging of living specimens that preserves physiology |
| Minimal interference from auto-fluorescence[3,4] | Irrelevant | Auto-fluorescence acquired as signal in label-free imaging (rather than background in fluorescence labeled imaging) |
| Aberration-free chromatic co-registration[5,6] | Irrelevant | Simultaneous multicolor signal detection at single-band excitation that guarantees this co-registration |
| Negligible out-of-focus background[3,4] | Irrelevant | Well-known strength of multiphoton microscopy that needs validation only at large imaging depths |
| Real-time monitoring of phototoxicity[4] | Irrelevant | Existence of an inline indicator in nonlinear optical imaging (elevated auto-fluorescence during time-lapse imaging) to monitor phototoxicity |
| Stable illumination[5,6] | Independent (Fig. 2, Fig. 3) | One constituent of PCEP required for Poisson noise-limited detection and flat-field illumination |
| Poisson noise-limited detection[5,6] | Independent (Fig. 2, Fig. 3) | One constituent of PCEP to enable *in situ* absolute measurement of photo-detection using a bulk fluorophore solution |
| Flat-field illumination[5,6] | Independent (Fig. 3, Suppl. Fig. 7) | One constituent of PCEP to enable uniform illumination across field-of-view of microscope objective |
| Accurate correction of color bleed-through or cross-talk[3,4] | Dependent (Fig. 4a,b, Suppl. Fig. 5) | One-time-calibrated photon crosstalk matrix to correct bleed-through in multicolor or dual-color ratiometric imaging, and the calibration is ensured by routine PCEP calibrations |
| Diffraction-limited lateral-axial resolution[5,6] | Dependent (Suppl. Fig. 10) | Standard one-time-calibrated point spread function ensured by routine PCEP calibrations |
| Real-time image processing and visualization[3,4] | Dependent (Suppl. Fig. 12) | Simple shading correction for different imaging modalities ensured by routine PCEP calibrations; real-time display free of image reconstruction |
| Repeatable stage positioning[5,6] | Feasible (Suppl. Fig. 13) | Optional calibration for large-field stitched imaging; not required for PCEP calibration and stitching-free imaging applications |
| Objective image quality and error assessment[3] | Feasible (Fig. 4b) | Quality assessment by representing each pixel as effective photons (with squared root as error according to Poisson statistics) |
| Standard image format and metadata for storage[3,4] | Feasible (Fig. 4b) | Universal pixel representation by effective photons (e.g., 0.001-2048 range); related metadata in photo-detection may be simplified |
| Real-time monitoring of photo-bleaching[3,4] | Feasible (Fig. 4c) | Rapid acquisition of 'noisy' images using low-power illumination or low-concentration fluorescence labeling |

Definition: Irrelevant – unnecessary subtasks if eSLAM is chosen for imaging; Independent – necessary subtasks for quality control by PCEP; Dependent – subtasks that only need one-time effort if PCEP calibration is routinely performed; Feasible – subtasks that are simplified by PCEP. Note: sample-dependent subtasks with no definitive measurable are highlighted in blue; technically challenging subtasks are highlighted in purple; independent subtasks are highlighted in red; other subtasks are highlighted in green.

**Supplementary Table 2.** Diverse elements of photo-detection (green) integrated in this study that are not directly related to microscopy quality control.

| Poisson-noise-limited detection (reference) | Photodetector | Sample | Illumination | Purpose |
|---|---|---|---|---|
| Varying-concentration[20] | *In situ* analog PMT vs PC PMT | NADH solutions | Two-photon | Advantage of PC over analog detection for weak signals |
| No[21] | PC PMT | Fluorophore solutions | Multi-photon | Absolute measurement of excitation molecular cross-section |
| Sample inhomogeneity[22] | *In situ* analog PMT vs PC PMT | FL-labeled polymer fibers | Two-photon | Advantage of PC over analog for weak signals via excess noise |
| Sample inhomogeneity[23] | *In situ* analog PMT vs SiPM | FL-labeled pollen grain | Two-photon | Advantage of siPM over analog PMT for strong signals via a SNR model |
| Varied frame number[24] | *In situ* analog PMT vs SiPM | Fluorescent test slide | Two-photon 1 pulse/pixel | Advantage of siPM over analog PMT for strong signals via PTC |
| Varying-power[25] | Diverse cameras | No | One-photon uniform light | Photon-resolving in camera-like detector via PTC |
| Varying-power[26] | Superconducting transition edge sensors | No | One-photon testing light | Photon-resolving in point-like detector by low read noise; quantum-information applications |
| Varying-power[27] | sCMOS camera | No | One-photon uniform light | Photon-resolving in camera-like detector by low read noise |

Note: FL – fluorescence; NADH – reduced nicotinamide adenine dinucleotide; PC – photon-counting; PMT – photomultiplier tube; PTC – photon transfer curve; sCMOS – scientific complementary metal-oxide-semiconductor; SiPM – silicon photomultiplier tube; SNR – signal to noise ratio.

**Supplementary Table 3.** Performance comparison among multiphoton microscopes in NADH/FAD imaging under safe illumination conditions (SLAM for theoretical comparison only).

| Microscope (reference) | SLAM (12) | **Traditional MPM (28,29)** | pSLAM (14) | **eSLAM** |
|---|---|---|---|---|
| Laser source | Custom fiber supercontinuum | Regular solid-state Ti:sapphire laser | Custom spectrally broadened laser | Custom fiber supercontinuum |
| Illumination band | 1110±30 nm | 750±5 nm (NADH) or 900±5 nm (FAD) | 1030±30 nm | 1110±30 nm |
| Pule repetition rate | 10 MHz | 80 MHz | 0.83 MHz | 5 MHz |
| Pulse width on sample | 60 fs (FWHM) | 200 fs (FWHM) | 60 fs (FWHM) | 60 fs (FWHM) |
| Water immersion microscope objective | UAPON40XW340 (Olympus), NA 1.15 | XLPLN25XWMP2 (Olympus), NA 1.05 | UAPON40XW340 (Olympus), NA 1.15 | UAPON40XW340 (Olympus), NA 1.15 |
| Illumination $P$ on sample* | 20 mW | 15 mW | 2.4 mW | 16.8 mW |
| Optical scanner | Galvo-Galvo | Galvo-Galvo | Galvo-Galvo | Resonant-Galvo |
| Fast scan line rate and readout mode | 140-350 Hz, unidirectional | 320 Hz, unidirectional | 340 Hz, bidirectional | 1592 Hz, bidirectional |
| Software control | LabVIEW | LabVIEW | ScanImage | LabVIEW |
| PMT – 2PF ($n$ = 2), quantum efficiency | H7421-40 (Hamamatsu), 31.6% | H7422P-40 (Hamamatsu), 42.1%, 43.2% | P2101 (Thorlabs), 40% | H7422A-40 (Hamamatsu), 41.4% |
| PMT – 3PF ($n$ = 3), quantum efficiency | H7421-40 (Hamamatsu), 31.8% | Not applicable | P2101 (Thorlabs), 44% | H7422A-40 (Hamamatsu), 42.1% |
| PMT – SHG ($n$ = 2), quantum efficiency | H7421-40 (Hamamatsu), 33.4% | Not applicable | P1001 (Thorlabs), 44% | H10721-20 (Hamamatsu), 16.8% |
| PMT – THG ($n$ = 3), quantum efficiency | H7421-40 (Hamamatsu), 20.4% | Not applicable | P2101 (Thorlabs), 24.2% | H10721-210 (Hamamatsu), 42.4% |
| Frame size (pixel × pixel) | 700 × 700 | 512 × 512 | 1024 × 1024 | 1024 × 1024 |
| Field-of-view (imaging) | 300 × 300 µm$^2$ | 128 × 128 µm$^2$ | 300 × 300 µm$^2$ | 250 × 250 µm$^2$ |
| Pixel dwelling time | 5-12 µs | 4 µs | 1.2 µs | 0.2 µs |
| pulses/pixel/frame ($m$) | 50-120 | 320 | 1 | 1 |
| Frame acquisition time | 2-5 s | 1.6 s | 1.5 s | 0.33 s |
| NADH and FAD imaging | Simultaneous | Sequential | Simultaneous | Simultaneous |
| Sensitivity: EP/pulse (10-mM NADH solution) | 0.21 | 0.27 | 0.35 | 1.0 |
| Throughput: EP/µs (10-mM NADH solution) | 2.1 | 22 | 0.29 | 5.1 |
| Sensitivity: EP/pulse (10-mM FAD solution) | 0.10 | 0.13 | 0.65 | 0.29 |
| Throughput: EP/µs (10-mM FAD solution) | 1.0 | 10 | 0.54 | 1.4 |

Note: EP – effective photon; * – illumination power is the upper limit tested on diverse live cell/tissue samples without observing the phototoxicity of elevated auto-fluorescence during time-lapse imaging.

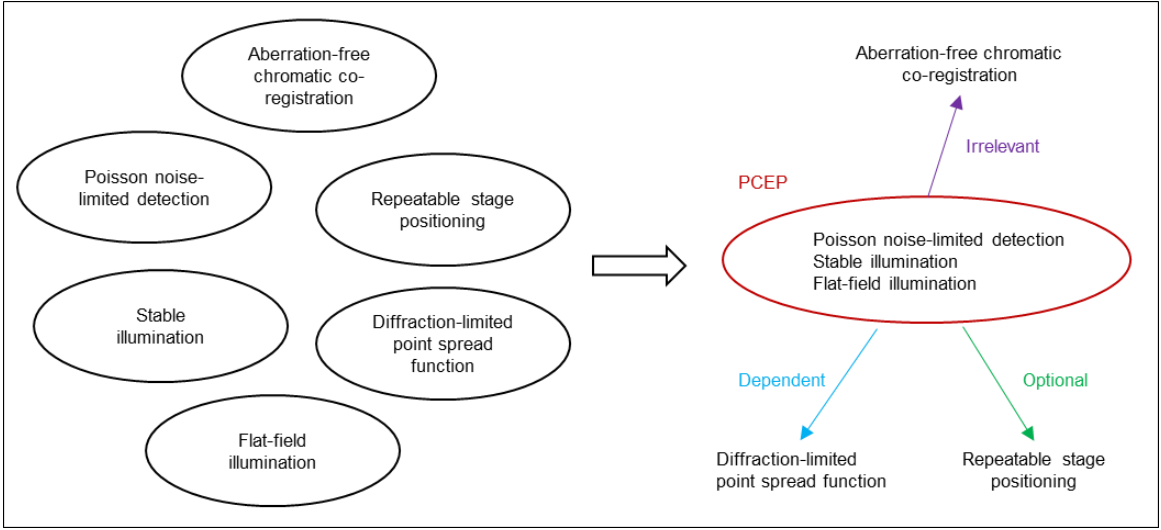

**Supplementary Fig. 1.** Reduction of 6 major quality-control subtasks to one routine procedure of PCEP.

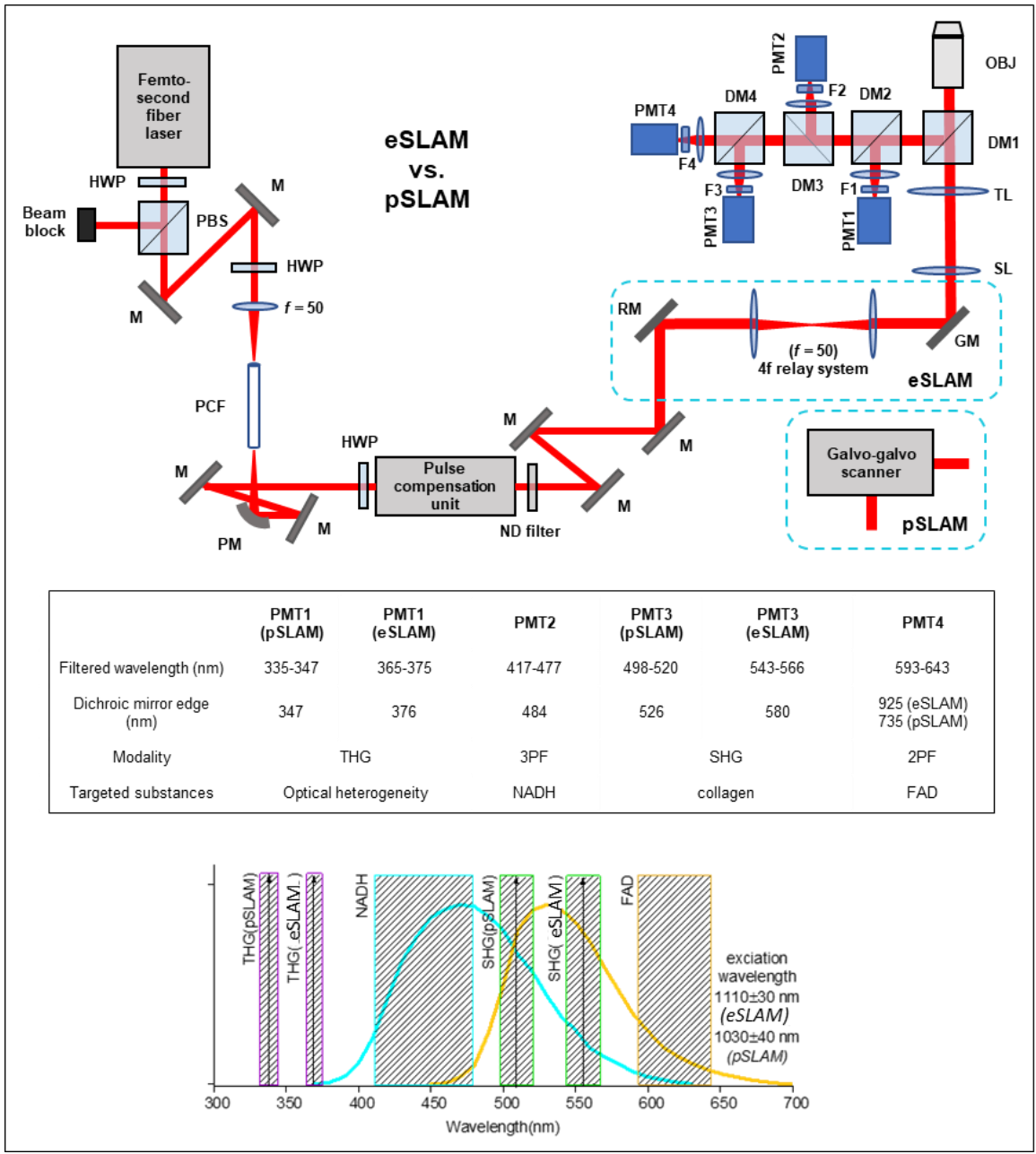

**Supplementary Fig. 2.** Detailed optical schematic of pSLAM and eSLAM (top) with 4 detection colors/channels (middle) that include NADH and FAD measurements (bottom). The inverted microscope consists of a nonlinear fiber spectrally broadened laser source with central wavelength of 1030-nm (pSLAM) or 1110-nm (eSLAM) and a pulse compensation unit (spatial light modulator-based pulse shaper for eSLAM and prism-based pulse compressor for pSLAM), while detection colors of THG, 3PF, SHG, and 2PF are spectrally separated according to the emission spectra of NADH and FAD (bottom). HWP – half wave plate; PBS – polarizing beam splitter; M – mirror; PCF – photonic crystal fiber; PM – parabolic mirror; HWP – half wave plate; ND – neutral density; RM – resonant mirror; GM – galvo mirror; SL – scan lens; TL – tube lens; DM – dichroic mirror; OBJ – objective; F – filter; PMT – photomultiplier tube.

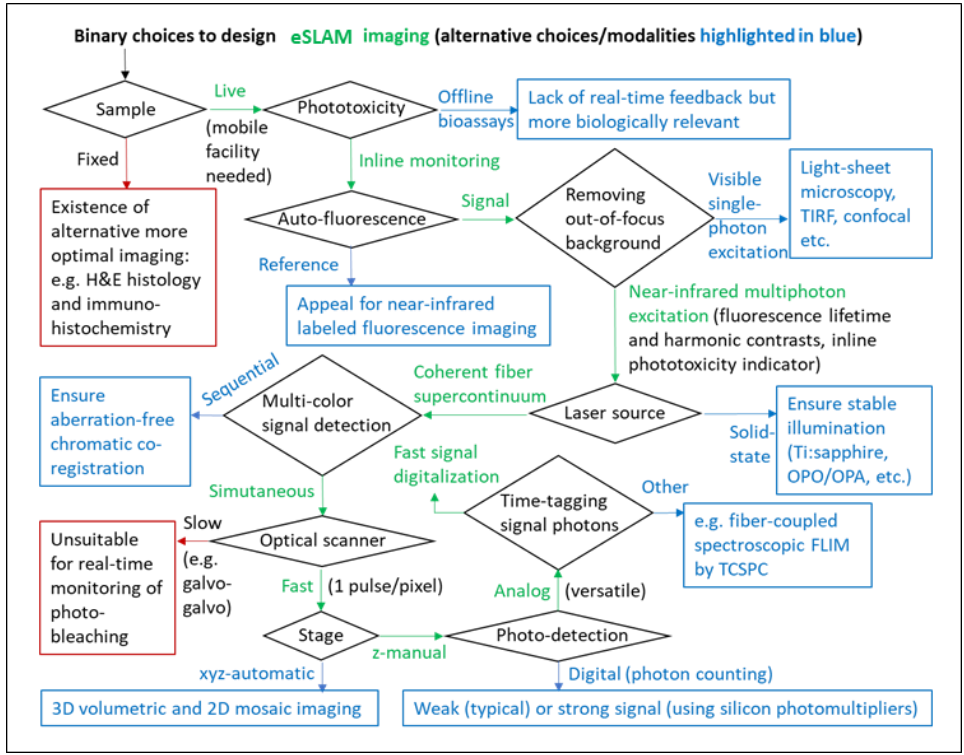

**Supplementary Fig. 3.** Diverse designs and binary choices of molecular optical sectioning microscopy alternative to eSLAM.

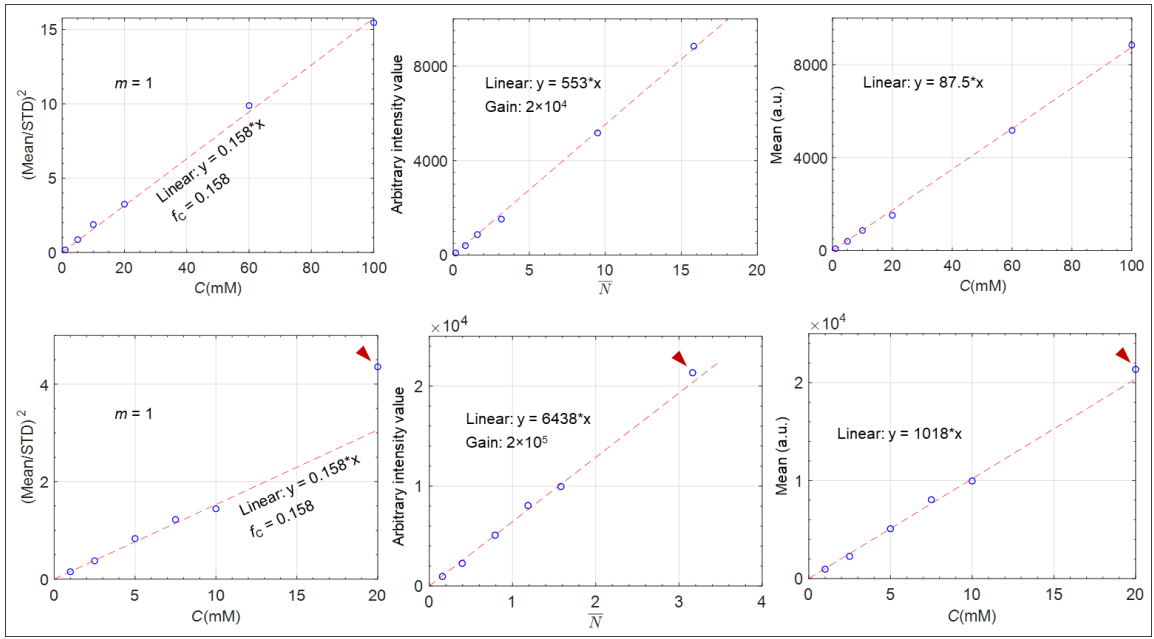

**Supplementary Fig. 4.** Alternative quantitative analysis of Fig. 2a. Left panels: determination of $f_C$ from single-parameter linear fit between experimental pSLAM (mean/STD)$^2$ and $C$ in 3PF "imaging" of 10 mM NADH solution with a small field-of-view at a low (top) and high gain (bottom); Middle panels: related conversion of arbitrary intensity values of PMT analog output to effective photons; Right panels: mean (in arbitrary intensity value) versus $C$ (in mM) that confirms a linear relation. Note that for figures associated with bottom panels at a high PMT gain, the (mean/STD)$^2$ vs. $C$ plot (lower left) is more sensitive to saturated photo-detection than the other two plots (arrowheads).

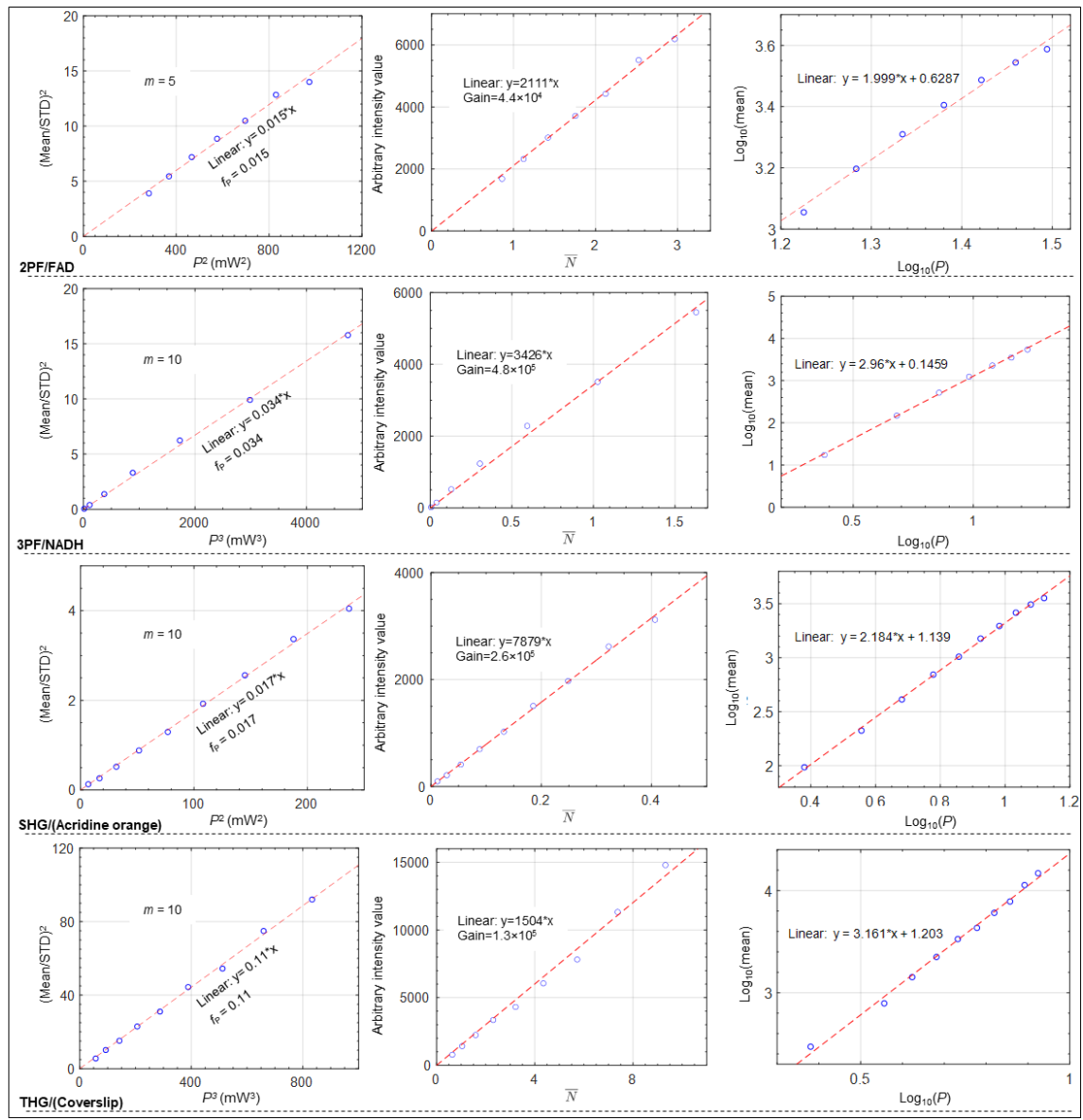

**Supplementary Fig. 5.** Quantitative analysis on multimodal/multicolor signals. Left panels: determination of $f_P$ from single-parameter linear fit between experimental eSLAM (mean/STD)$^2$ and $P^n$ in 2PF, 3PF, SHG, and THG "imaging" of a small field-of-view, which involves 50-mM FAD (2PF), 10-mM NADH solutions (3PF), 1 mg/mL acridine orange solution that mimics a homogeneous SHG sample (SHG), and coverslip surface (THG), respectively; Middle panels: related conversion of arbitrary intensity values of PMT analog output to effective photons corresponding to the 4 imaging modalities at specific PMT gains; Right panels: mean (in arbitrary intensity value) versus $P$ (in mW) in the log-scale that is consistent with the nonlinear photon order of 2 (2PF, SHG) or 3 (3PF, THG).

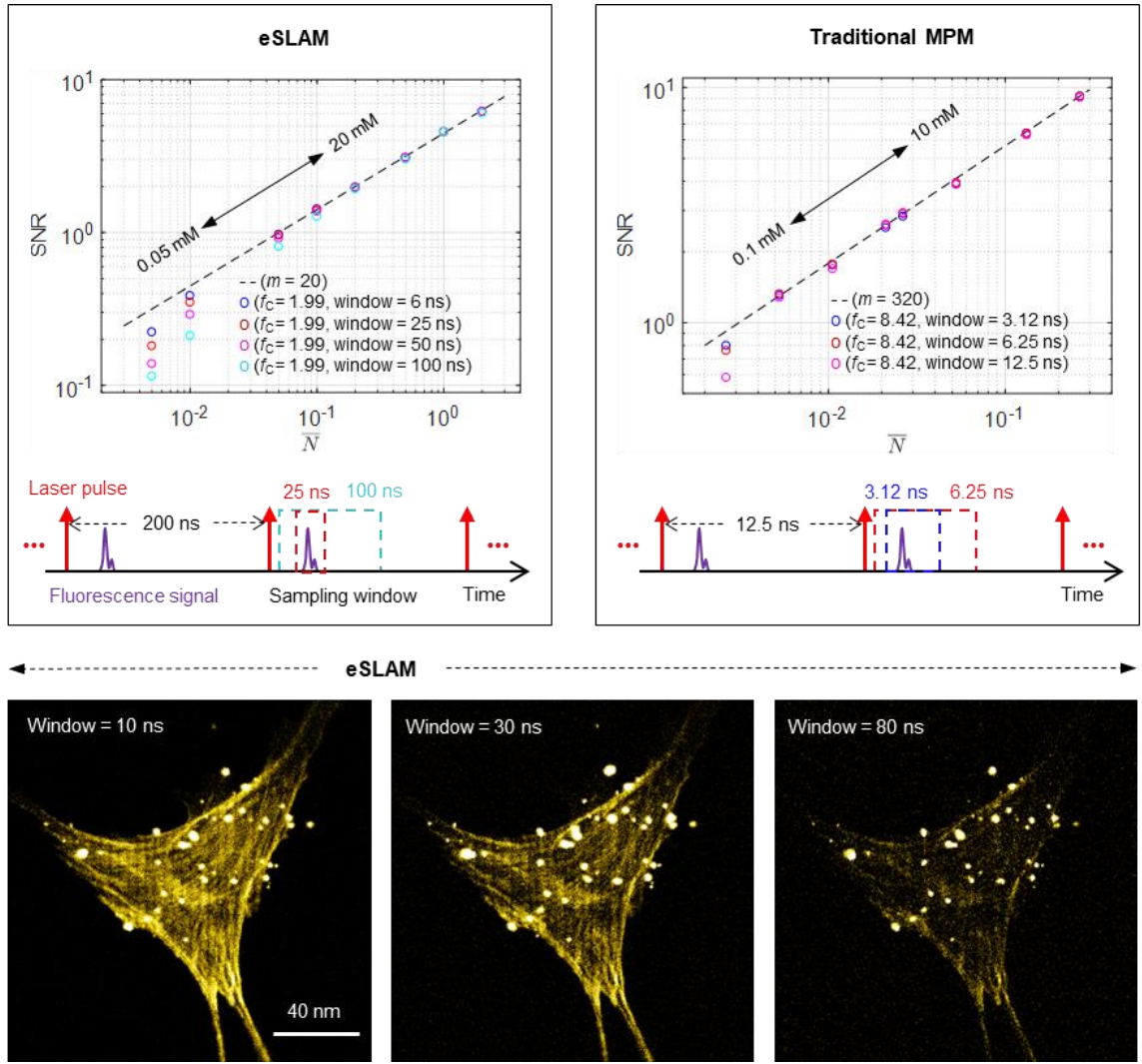

**Supplementary Fig. 6.** Optimization of temporal window of signal collection to obtain the largest Poisson-noise-dominated dynamic range (PDR) without loss of signal. (Top) Dependence of low limit of PDR on temporal window in eSLAM with 200-ns pulse separation (upper left) and a traditional multiphoton microscope MPM with 12.5-ns pulse separation (upper right); (Bottom) related effect on eSLAM (2PF) imaging of fixed fluorescent cells (ThermoFisher) showing the ability of optimized temporal window to detect weak signal.

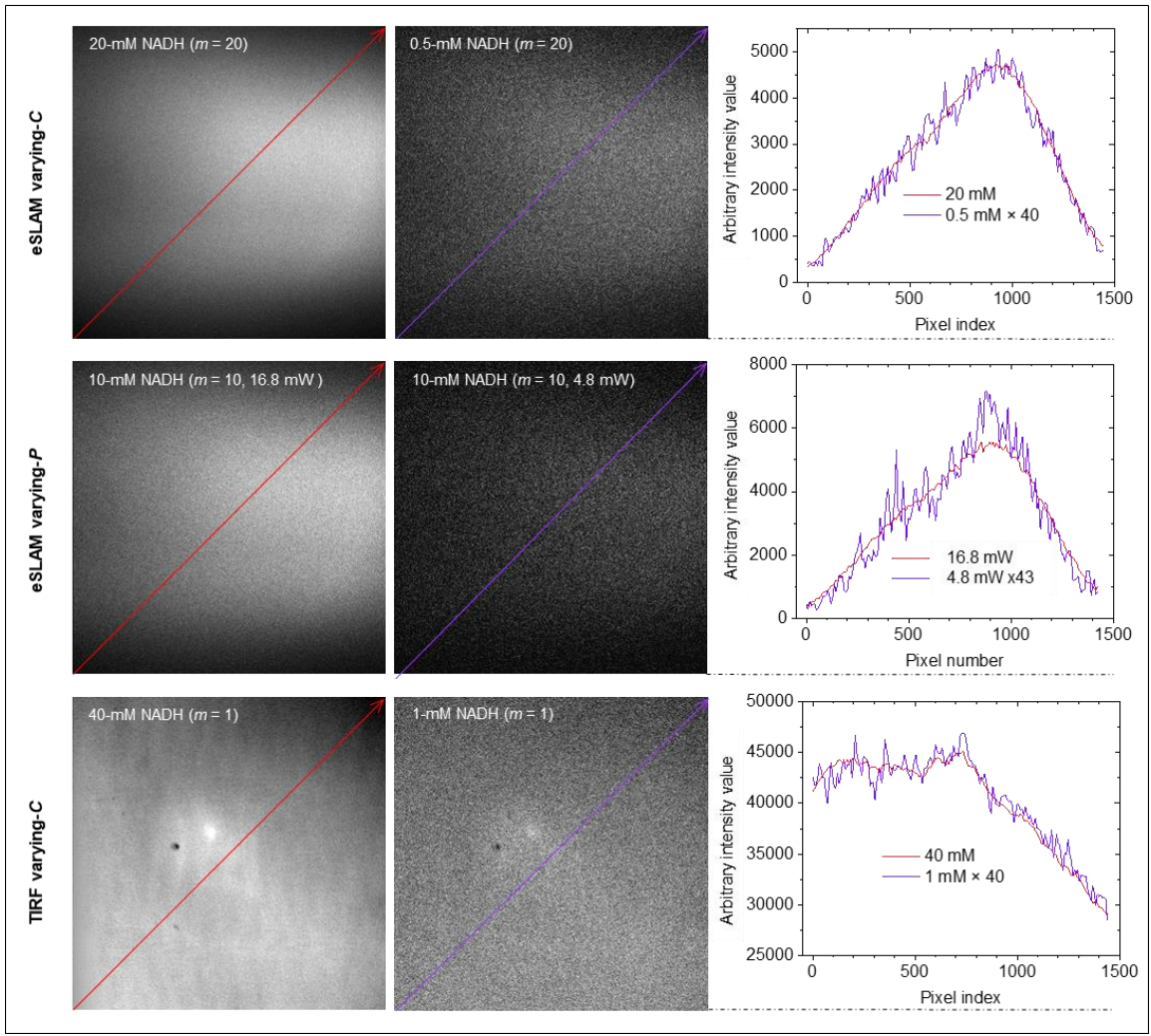

**Supplementary Fig. 7.** Visualized illumination field of eSLAM (top, middle) or TIRF (bottom) in a bulk solution independent of NADH concentration (top, bottom) and illumination power (middle).

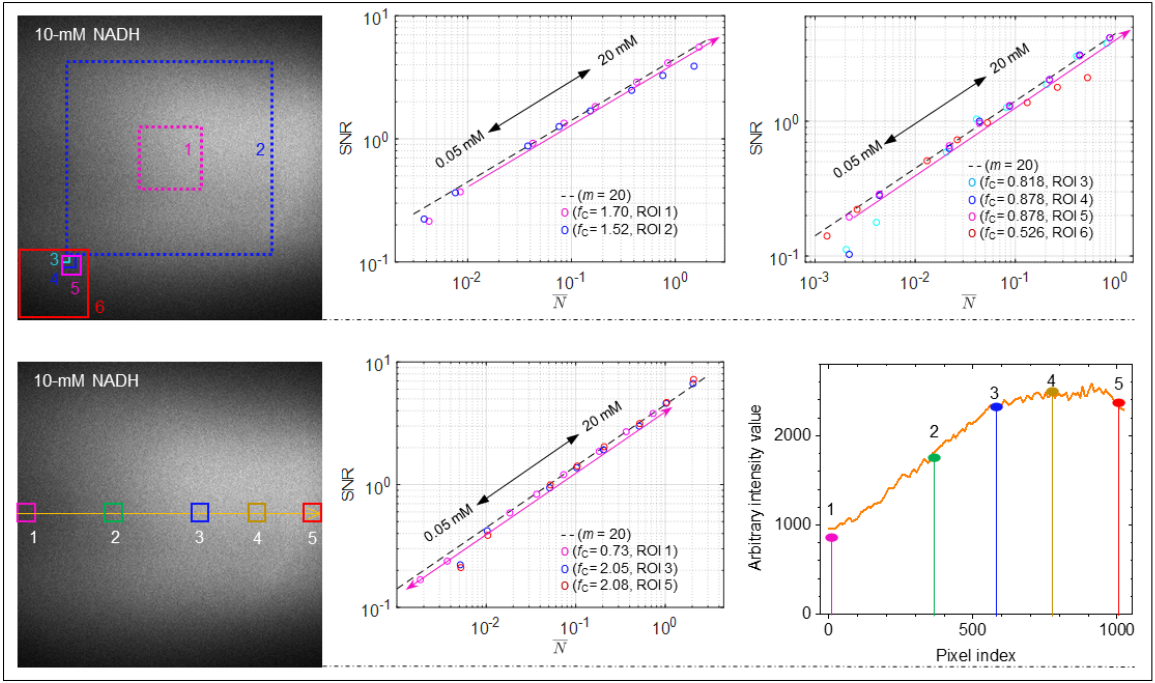

**Supplementary Fig. 8.** Detailed PCEP calibration at $P$ = 16.8 mW. (Top) Dependence of observed Poisson-noise-dominated dynamic range (PDR) on the size and location of region-of-interest (ROI) inside the illumination field. (Bottom) Convergent PDRs of different super-pixels with $f_C$ parameters proportional to local illumination field strengths (right panel).

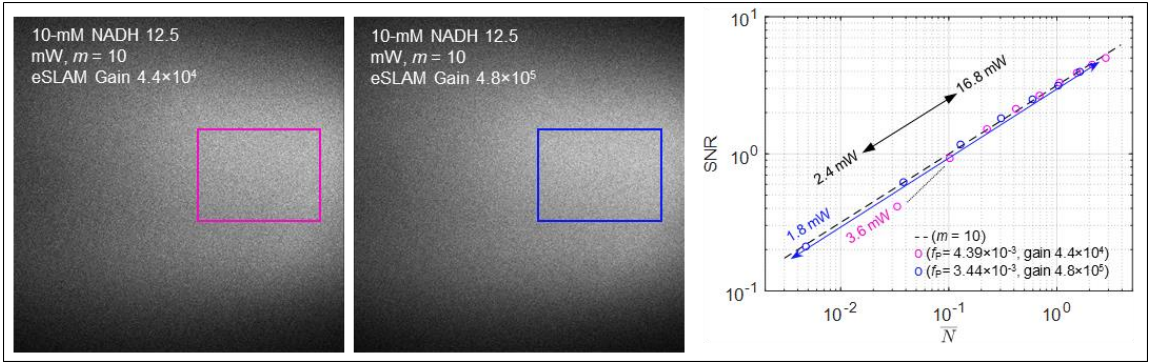

**Supplementary Fig. 9.** Decreased detection sensitivity and PDR at a lower PMT gain in the eSLAM varying-$P$ experiment of Fig. 3f.

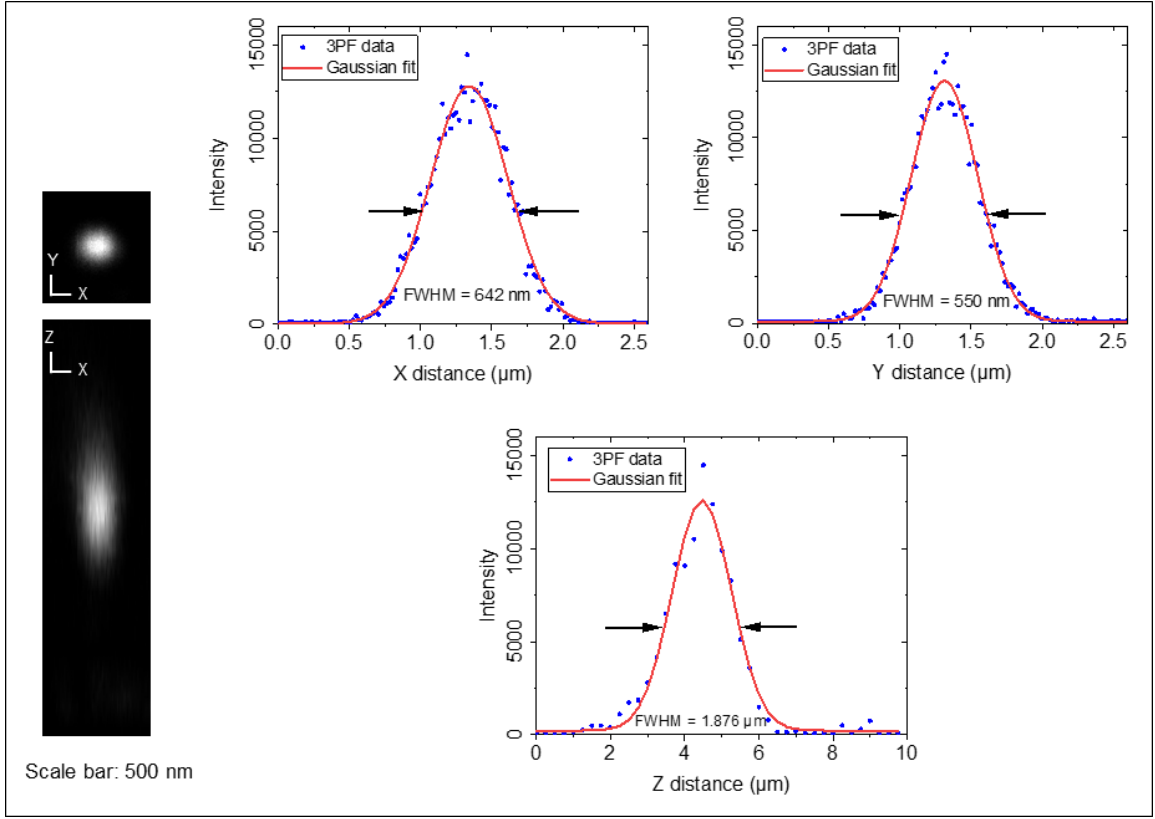

**Supplementary Fig. 10.** Measurement of point spread function of eSLAM via a fluorescent bead following similar measurement (Biomed. Opt. Express 13:452, 2022). Fluorescent beads of 100 nm diameter (Fluorophorex polystyrene nanospheres, No. 2002 with EM/EX 345/435 nm Phosphorex, Inc) were embedded in an agarose solution and used for the PSF measurements of three-photon fluorescence imaging. The resulting 3D images were collected over a FOV of 20 μm × 20 μm with 0.25 μm axial step and 10 μm total depth.

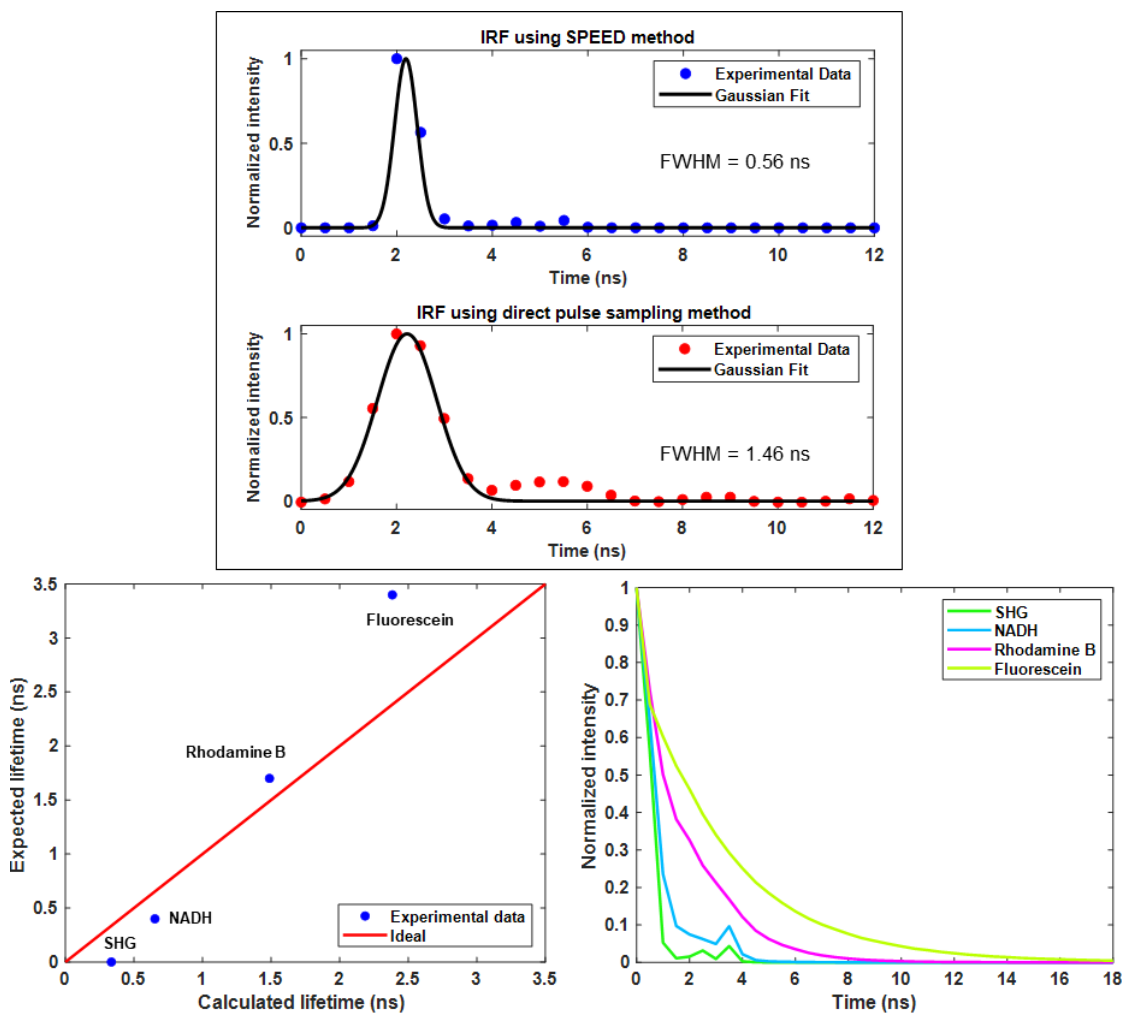

**Supplementary Fig. 11.** (Top) Impulse response functions (IRF) in eSLAM based on SHG of urea crystal using SPEED method and the direct pulse sampling method, with Gaussian fits to estimate the full-width-half-maximum in ns; (Bottom) Calculated lifetimes and expected values for four different fluorophores/harmonophores (left) and fluorescence decay curves for each calibration fluorophore/harmonophore (right). For short lifetimes, such as NADH and SHG, a reflection from the amplifier may be present a few ns after photon arrival and may register as a photon count.

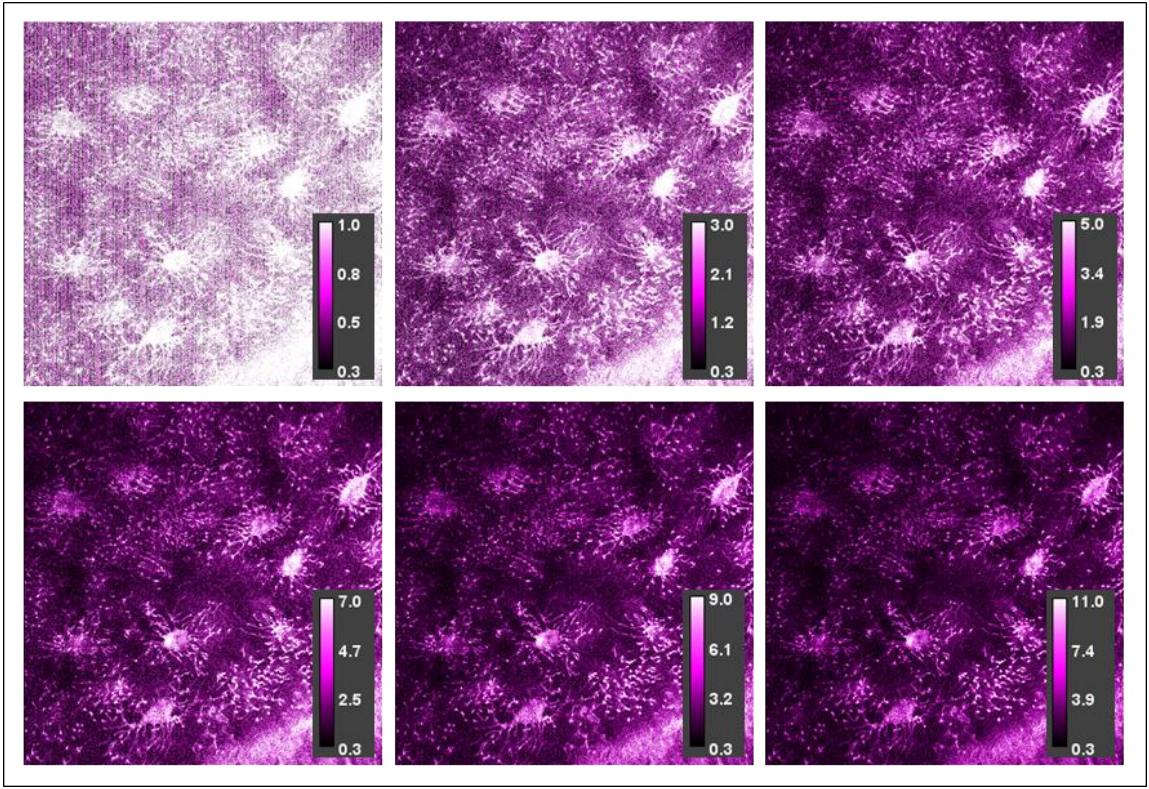

**Supplementary Fig. 12.** Single-frame THG images of osteocytes in *ex vivo* mouse skull from gentle eSLAM imaging (13 mW on sample). The high dynamic range of the images is displayed with different scales in effective photons while the statured image (upper left) associated with scale (0-1) is expected from the photon-counting detection of SLAM.

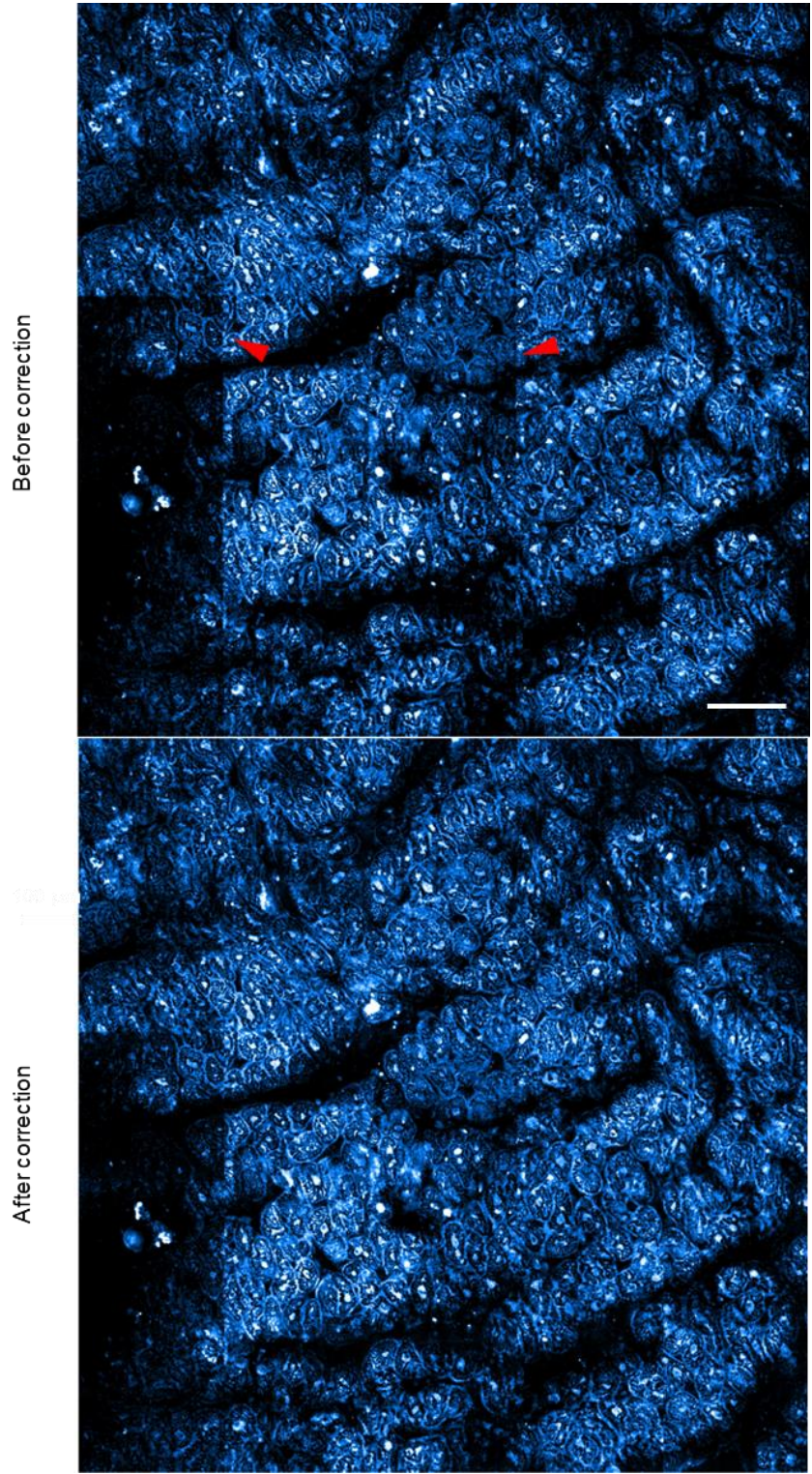

**Supplementary Fig. 13.** Shading correction of effective photon-pixelated 5×5 mosaic-stitched 3PF image from *ex vivo* rabbit heart tissue. While a large overlapping ratio for adjacent field-of-views (~30%) limits shading effect (top), residual artifacts (arrowheads) can be largely corrected by the known 3PF field from an NADH solution (bottom). Scale bar: 100 µm.

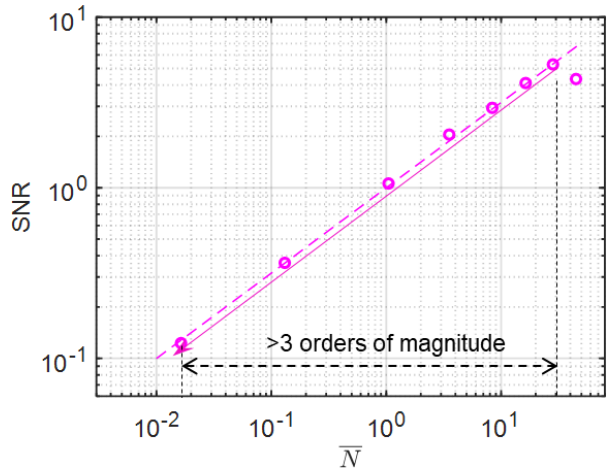

**Supplementary Fig. 14.** High dynamic range PDR with >3 orders of magnitudes in effective photons in the experiment of Fig. 2c (THG in pSLAM).

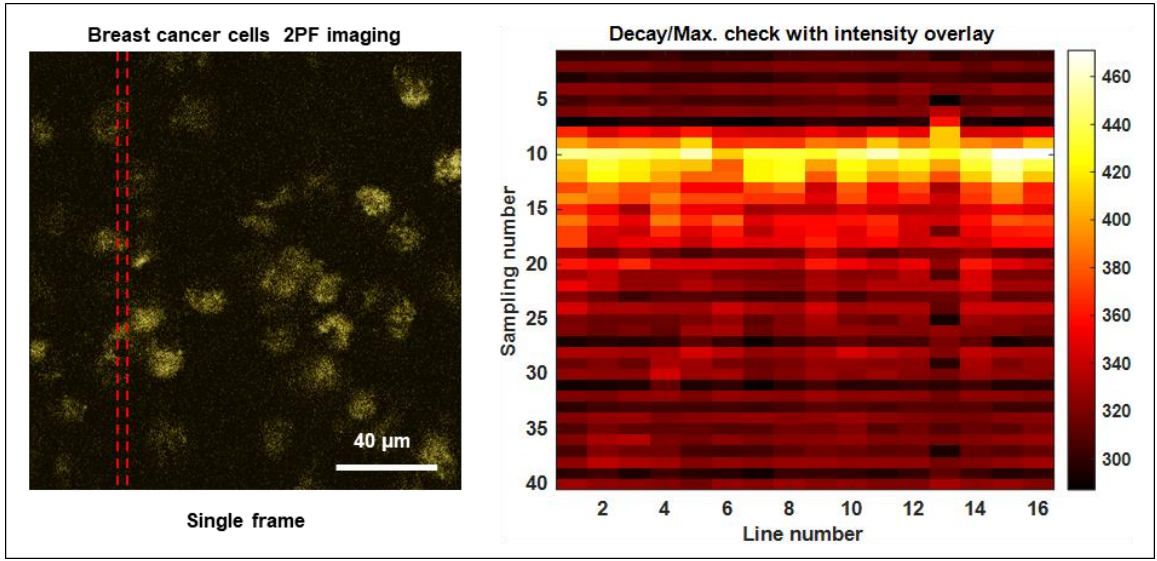

**Supplementary Fig. 15.** Demonstration of time-gated window with 40 sampling points. (Left) Single-frame 2PF image of unlabeled breast cancer cells with a narrow red dotted area of interest. (Right) Superimposition of all the pixels within each line of this area of interest reveals a time-gated window with 40 sampling points, in which the maximum value is set at the 10$^{th}$ sampling point.

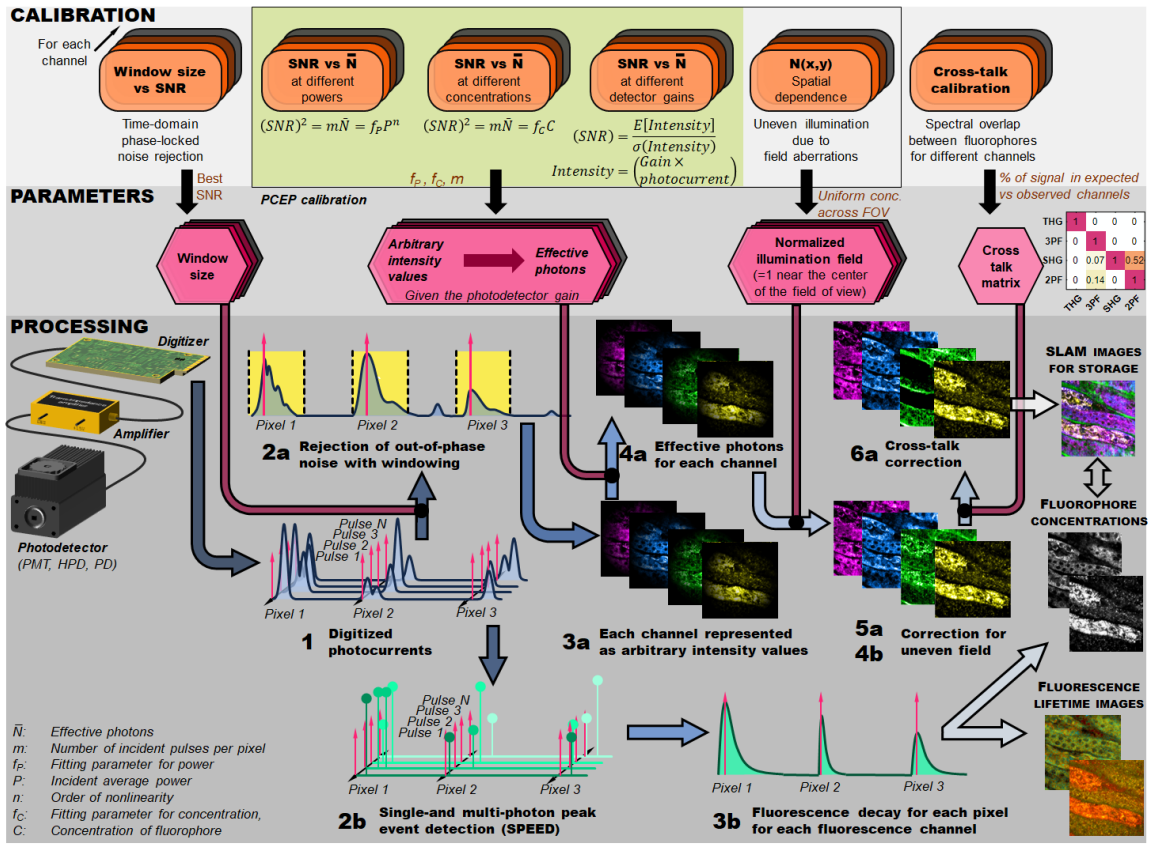

**Supplementary Fig. 16.** Overview of PCEP calibration and image processing in eSLAM.

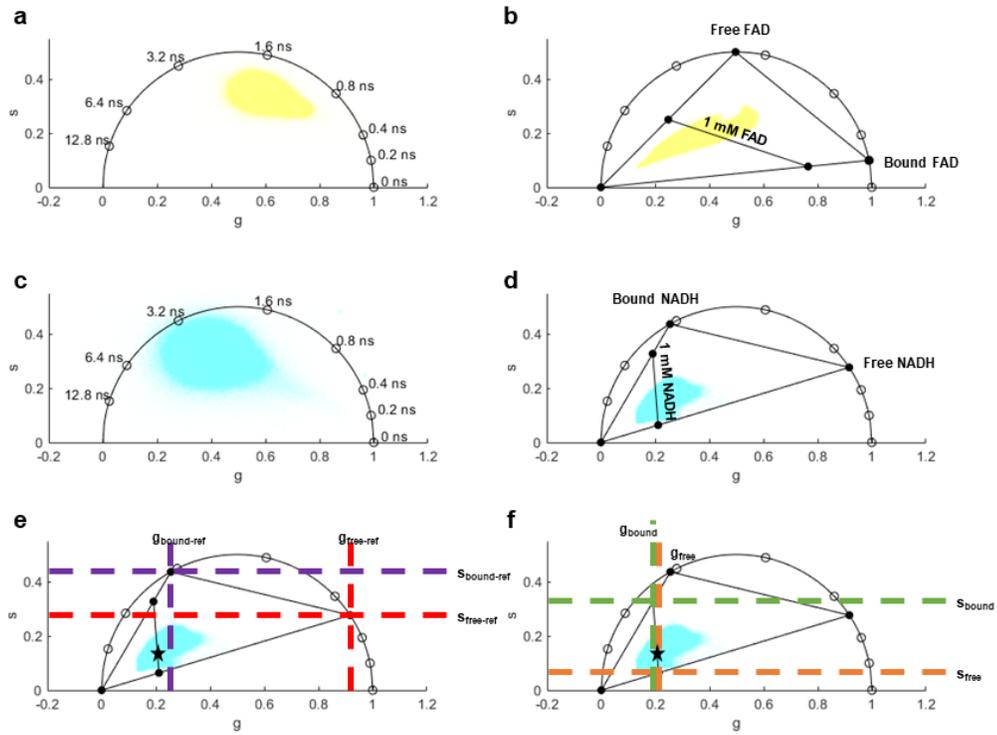

**Supplementary Fig. 17.** Determination of FAD and NADH concentration using phasor plots.